\documentclass[aps, prd, reprint, amsfonts,
  amssymb, amsmath, showpacs, preprintnumbers, letterpaper,
  nofootinbib,floatfix,superscriptaddress]{revtex4-1}

\usepackage{braket,bm}
\usepackage[pass]{geometry} 
\usepackage{color}
\usepackage{mathrsfs,mathtools}
  \DeclareMathAlphabet{\mathpzc}{OT1}{pzc}{m}{it}
  
\newcommand{\D}{\mathrm{d}}
\renewcommand{\|}{\big|\hspace{-1pt}\big|}

\begin{document}

%%%%%%%%%%%%%%%%%%%%%%%%%%%%%%%%%%%%%%%%%%%%%%%%%%%%%%%%%%%%%%%%%%%%%%

\title{Self-consistent solitons for vacuum decay in radiatively generated potentials}

\author{Bj\"{o}rn Garbrecht}
\email{garbrecht@tum.de}
\affiliation{Physik Department T70, Technische Universit\"{a}t M\"{u}nchen,\\ James-Franck-Stra\ss e, 85748 Garching, Germany}

\author{Peter Millington}
\email{p.millington@nottingham.ac.uk}
\affiliation{Physik Department T70, Technische Universit\"{a}t M\"{u}nchen,\\ James-Franck-Stra\ss e, 85748 Garching, Germany}
\affiliation{School of Physics and Astronomy, University of Nottingham,\\
Nottingham NG7 2RD, United Kingdom}

% quantum field theory (first two), quantum tunneling of defects

\preprint{TUM-HEP-1017-15}

%%%%%%%%%%%%%%%%%%%%%%%%%%%%%%%%%%%%%%%%%%%%%%%%%%%%%%%%%%%%%%%%%%%%%%
\begin{abstract}
We use a Green's function approach in order to develop a method for calculating the tunneling rate between radiatively-generated non-degenerate vacua. We apply this to a model that exhibits spontaneous symmetry breaking via the Coleman-Weinberg mechanism, where we determine the self-consistent tunneling configuration and illustrate the impact of gradient effects that arise from accounting for the underlying space-time inhomogeneity.
\end{abstract}

\maketitle

%%%%%%%%%%%%%%%%%%%%%%%%%%%%%%%%%%%%%%%%%%%%%%%%%%%%%%%%%%%%%%%%%%%%%%

\section{Introduction}

State-of-the-art calculations~\cite{EliasMiro:2011aa,Degrassi:2012ry,Alekhin:2012py, Buttazzo:2013uya,Bednyakov:2015sca} suggest that the electroweak vacuum of the Standard Model suffers an instability at a scale of around $10^{11}$ GeV, with the lifetime of the electroweak vacuum lying in the so-called metastable region and being longer than the current age of the Universe (for a recent overview, see Ref.~\cite{DiLuzio:2015iua}). The origin of this instability is the generation of a high-scale global minimum in the Higgs potential through radiative effects due to the renormalization-group running of the Higgs quartic self-coupling~\cite{Cabibbo:1979ay,Sher:1988mj,Sher:1993mf,Isidori:2001bm}. Specifically, when one applies the standard renormalization procedure, this coupling is driven negative by top-quark loops, with the dominant experimental uncertainty originating from the current measurement of the top mass~\cite{Bezrukov:2012sa,Masina:2012tz}. The latter effect has, however, been challenged recently~\cite{Gies:2014xha} in the light of contradictory observations from lattice simulations of Higgs-Yukawa models~\cite{Holland:2003jr,Holland:2004sd,Fodor:2007fn}, where the full effective potential is found to remain stable so long as the ultraviolet cutoff is kept finite. Moreover, the presence of new physics at high scales has been shown to have a dramatic impact upon the tunneling rate~\cite{Branchina:2013jra, Branchina:2014usa, Branchina:2014rva, Lalak, Eichhorn:2015kea,Branchina:2015nda}.

Often, the tunneling rate is determined from the effective potential~\cite{Jackiw:1974cv,Cornwall:1974vz} calculated assuming a homogeneous field configuration, which is subsequently promoted to an inhomogeneous field configuration~\cite{Frampton:1976kf,Frampton:1976pb}. Thus, the impact of the space-time dependence of the underlying tunneling configuration is not fully accounted for.

In light of the aforementioned theoretical and phenomenological questions, it is timely to consider approaches to the calculation of tunneling rates from false vacua that can consistently account for radiative effects in the inhomogeneous solitonic background of the tunneling configuration. This is all the more relevant when the global minimum of the potential emerges entirely through radiative effects. In this article, we apply the Green's function method developed in Ref.~\cite{Garbrecht:2015oea} to the calculation of the one-loop tunneling rate in a model with spontaneous symmetry breaking (SSB) that arises purely radiatively via the Coleman-Weinberg mechanism~\cite{Coleman:1973jx}. Green's function methods have previously been applied to determine self-consistent bounce solutions in the Hartree-Fock approximation of the pure $\lambda\Phi^4$ theory~\cite{Bergner:2003au,Bergner:2003id,Baacke:2004xk,Baacke:2006kv}. This article represents a first exercise in the use of the Green's function method in Ref.~\cite{Garbrecht:2015oea} for dealing with potentials that are \emph{significantly} modified by radiative effects, the aim being to understand the parametric dependencies of the tunneling rate and the relative importance of gradient effects. The latter effects are anticipated to be small, contributing corrections at an order comparable to two-loop effects~\cite{Weinberg:1992ds}, and we present herein a numerical confirmation of this observation.

In the present analysis, we will consider the importance of accounting for the aforementioned space-time dependence of the tunneling configuration in the case of \emph{spontaneous} decay of an initially \emph{homogeneous} false vacuum state. This is in contrast to \emph{induced} transitions where
an \emph{inhomogeneous} initial state acts as a potential seed for vacuum decay, as has been studied for the case of black holes~\cite{Burda:2015yfa}, for topological defects~\cite{Kumar:2010mv,Lee:2013ega,Lee:2013zca,Dupuis:2015fza,Haberichter:2015xga} and, in the context of the Standard Model, for impurities in the Higgs vacuum \cite{Grinstein:2015jda}. Such seeds may lead to an enhanced decay rate.

The remainder of this article is organized as follows. In Sec.~\ref{sec:model}, we describe the renormalized one-loop effective potential of the model under consideration. Additional technical details are provided in Appendix~\ref{app:CW}. In Sec.~\ref{sec:tunnel}, we outline the calculation of the tunneling rate in this model, making comparisons with the equivalent calculation in the case of tree-level vacuum instability. Details of the method used to calculate the fluctuation determinant are given in Appendix~\ref{app:Baacke}. In Sec.~\ref{sec:numproc}, we describe the numerical procedure employed for determining the self-consistent tunneling configuration, the results of which are presented in Sec.~\ref{sec:numres}. Our conclusions and potential future directions are given in Sec.~\ref{sec:conc}.

\section{Model}
\label{sec:model}

We consider a scalar model with the following Euclidean Lagrangian density:
\begin{equation}
\label{eq:lagrangian}
\mathcal{L}  = \frac{1}{2}\,\big(\partial_{\mu}\Phi\big)^2+\frac{1}{2}\sum_{i\,=\,1}^N\big(\partial_{\mu}X_i\big)^2 +U\;,
\end{equation}
with the tree-level potential
\begin{equation}
\label{eq:potential}
U =  \frac{\lambda}{4}\,\Phi^2\sum^N_{i\,=\,1}X_i^2+\frac{\kappa}{4}\sum_{i,j\,=\,1}^NX_i^2X_j^2+\frac{g}{3!}\,\Phi^3 +U_0\;,
\end{equation}
comprising a real scalar field $\Phi_x\equiv\Phi(x)$ and $N$ real scalar fields $X_{i,x}\equiv X_i(x)$, $i=1,2,\dots,N$. The small cubic coupling $g$, of mass dimension 1, has been added by hand to break the $\mathbb{Z}_2$ symmetry, and $U_0$ is a constant.

\begin{figure}
\includegraphics[scale=0.6]{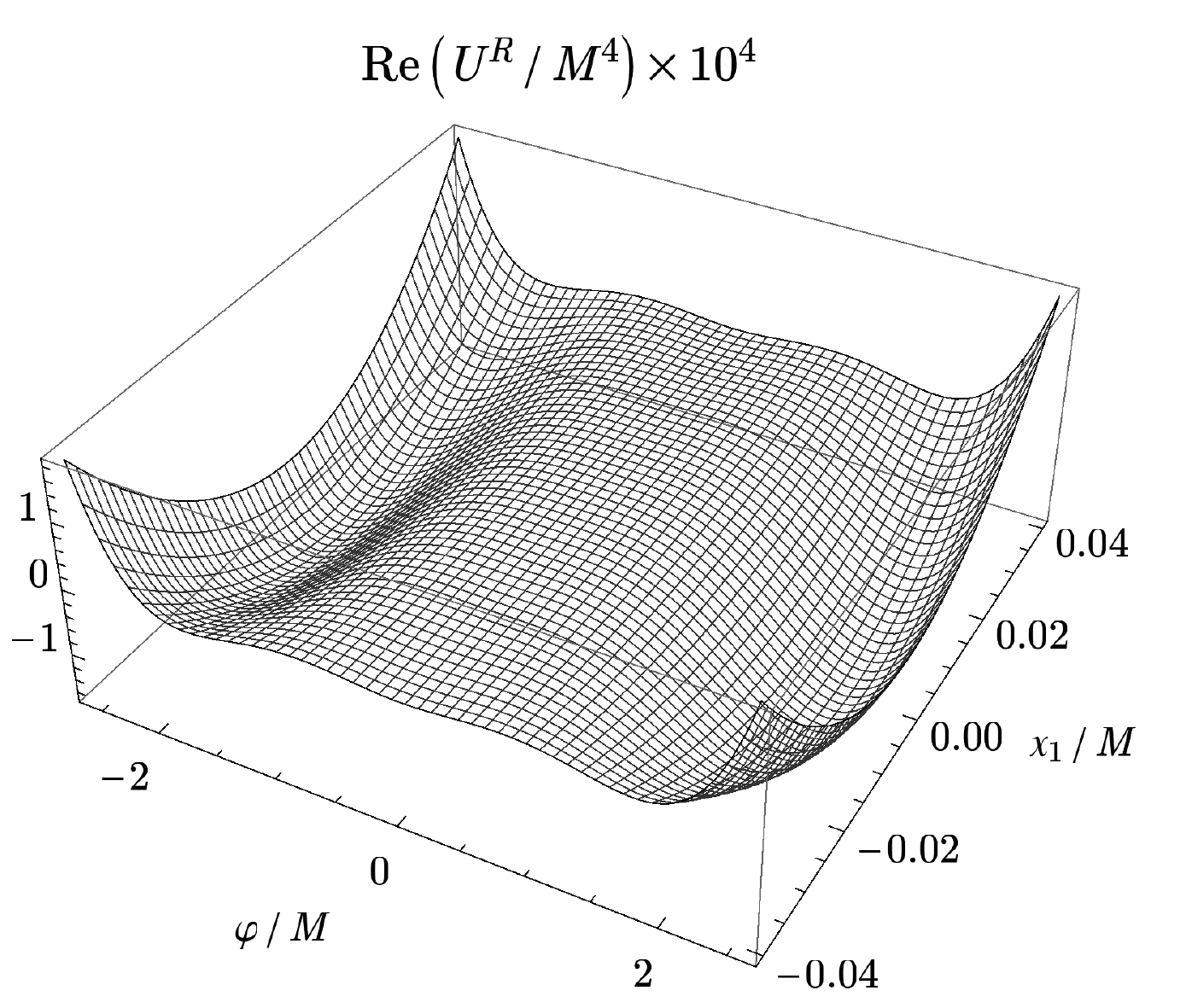}
\caption{\label{fig:3d}The real part of the renormalized effective potential in the unitary gauge ($\chi_i=0$, $i\neq 1$), plotted as a function of the vacuum expectation values $\varphi$ and $\chi_1$ for the parameters $g\to0$, $\lambda=0.1$, $\kappa=0.05$, $M=1$ and $N=4$.}
\end{figure}

\begin{figure}
\includegraphics[scale=0.6]{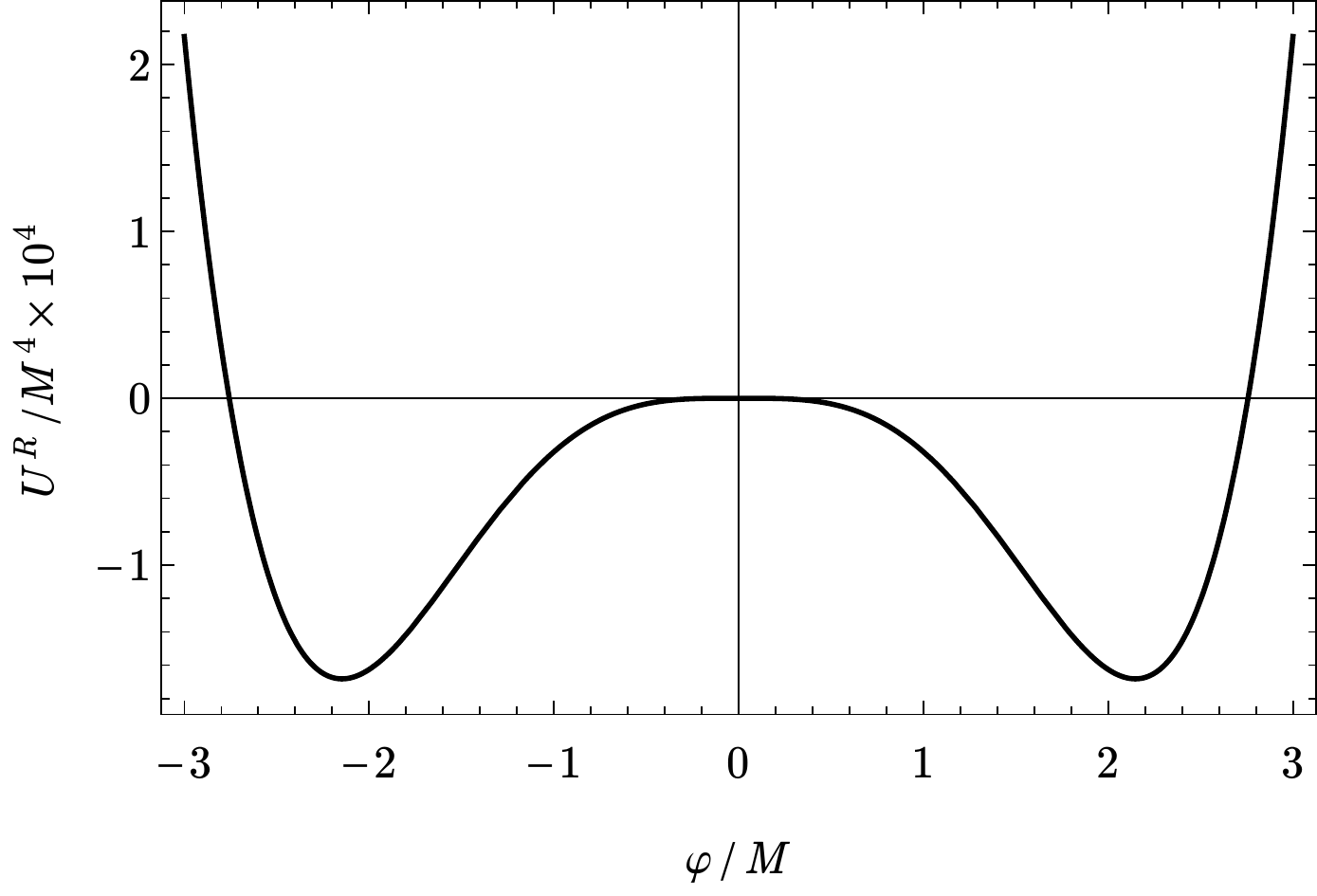}
\caption{\label{fig:1d}The effective potential in the unitary gauge, plotted as a function of the vacuum expectation value $\varphi$ at $\chi_1=0$ and for the parameters $g\to0$, $\lambda=0.1$, $\kappa=0.05$, $M=1$ and $N=4$.}
\end{figure}

The classical scale invariance of this model (which is present in the limit $g\to 0$) is broken by radiative effects at the one-loop level. 
Logarithmic infrared divergences require the introduction of a dimensionful renormalization scale $M$ that is turned into a symmetry-breaking scale by the well-known mechanism of dimensional transmutation~\cite{Coleman:1973jx}.

The field $\Phi$ develops a vacuum expectation value $\varphi=\braket{\Phi}$ at $\chi_i=\braket{X_i}=0$. In the direction $\chi_i=\braket{X_i}=0$, the renormalized effective potential of the homogeneous field configuration (see Appendix~\ref{app:CW}) is given by
\begin{align}
\label{eq:Ueffrho}
U_{\mathrm{eff}}^R(\varphi)\ &=\  \bigg(\frac{\lambda}{16\pi}\bigg)^{\!2}\varphi^4\bigg[N\bigg(\!\ln\frac{3\varphi^2}{\rho M^2}\:-\:\frac{3}{2}\bigg)\:-\:F\bigg]\nonumber\\& \qquad +\:\frac{g}{6}\,\varphi^3\:+\:U_0\;.
\end{align}
Here, $\rho=6\kappa/\lambda$ is the ratio of the couplings,
\begin{equation}
F\ \equiv\ \ln 3\:+\:\frac{8}{(1-\rho)^2}\bigg(3\:+\:\rho\:+\:\frac{1+3\rho}{1-\rho}\,\ln \rho\bigg)\;,
\end{equation}
and we have neglected terms of order $g^2$. Note that the limit $\rho\to 1$ ($\lambda\to 6\kappa$) is well defined, with $F\to\ln3+4/3$. For $g\ll 32\pi^2 v/(\lambda^2N)$, the minima of the effective potential lie at
\begin{equation}
\label{eq:vev}
\varphi\ \approx\ \pm\,v\ =\ \pm\,\sqrt{\frac{\rho M^2}{3}}\,\exp\bigg[\frac{1}{2}\:+\:\frac{F}{2N}\bigg]\;,
\end{equation}
depending only on the ratio $\rho$ of the couplings. The constant shift in the potential is fixed to be $U_0 = -\,gv^3/6$, such that the false vacuum has zero energy density.

We note from Eq.~\eqref{eq:vev} that the value of $v$ is of order $M$ as long as $\rho$ remains of order one. Consequently, the logarithm in Eq.~\eqref{eq:Ueffrho} will be of order one as well. This is in contrast to the well-known case of a single field with quartic self-interactions, where the corresponding logarithm turns out to be enhanced by an inverse power of the coupling constant thereby invalidating perturbation theory~\cite{Coleman:1973jx}. Of course, the model specified in Eqs.~\eqref{eq:lagrangian} and~\eqref{eq:potential} represents only one of many possibilities of implementing radiative symmetry breaking in a perturbatively self-consistent manner.

In Fig.~\ref{fig:3d} and in order to illustrate that the global minima of the potential do indeed lie at $\chi_i=\braket{X_i}=0$, we plot the real part of the effective potential in the unitary gauge ($\chi_i=0$, $i\neq 1$) for $g\to0$, $\lambda=0.1$, $\kappa=0.05$, $M=1$ and $N=4$. There also exist two shallow local minima along $\varphi=0$, within which the field $\chi_1$ develops a non-zero vacuum expectation value. In addition, we note that, for $\chi>\varphi$ and $\varphi\neq 0$, the effective potential acquires an imaginary part as a result of one of the mass eigenvalues (see Appendix~\ref{app:CW}) becoming tachyonic. Nevertheless, at $\chi_1=0$, the effective potential is real valued and takes the form shown in Fig.~\ref{fig:1d}. We will not discuss the additional local minima or the tachyonic modes further, since they are of little relevance to the forthcoming analysis.

Hereafter, we will fix the ratio $\kappa/\lambda=1/2$, in which case $\rho = 3$. The effective potential then takes the form
\begin{equation}
U_{\mathrm{eff}}^R(\varphi) =  \frac{\lambda^2}{16^2\pi^2}\,\varphi^4\bigg[N\bigg(\!\ln\frac{\varphi^2}{M^2}-\frac{3}{2}\bigg)-F\bigg]\;,
\end{equation}
with the constant $F$ simplifying to
\begin{equation}
F\ =\ 3\big(4\:-\:3\ln\,3\big)\ \approx\ 2\;.
\end{equation}
The minima of the potential now lie at
\begin{equation}
\pm\,v\ =\ \pm\,M\,e^{\frac{1}{2}+\frac{F}{2N}}\ \approx\ \pm\,M\,e^{\frac{1}{2}}\;,
\end{equation}
where the approximation holds for $N\gg 2$. Thus, for large $N$, the positions of the minima are constant with respect to both the couplings and the number of $X$ fields.

\section{Tunneling rate}
\label{sec:tunnel}

Before proceeding to discuss the scale-invariant model described in the preceding section, it is helpful first to review the most pertinent details of the calculation of the tunneling rate in a model that exhibits non-degenerate vacua at tree level. The archetypal example is the $\lambda\Phi^4$ theory with tachyonic mass term (see e.g.~Refs.~\cite{Coleman:1977py,Callan:1977pt}), having the Lagrangian
\begin{equation}
\mathcal{L}\ =\ \frac{1}{2}\,\big(\partial_{\mu}\Phi\big)^2\:-\:\frac{1}{2}\,\mu^2\Phi^2\:+\:\frac{\lambda}{4!}\,\Phi^4\:+\:\frac{g}{3!}\,\Phi^3\:+\:U_0\;,
\end{equation}
where $\mu^2>0$. For $g^2/\mu^2\ll 8\lambda/3$, the potential has non-degenerate minima at
\begin{equation}
\varphi\ \approx\ \pm\,v\ =\ \pm\,\sqrt{\frac{6\mu^2}{\lambda}}\;.
\end{equation}
The classical equation of motion for this theory is
\begin{equation}
\frac{\delta S[\Phi]}{\delta \Phi_x}\bigg|_{\Phi\,=\,\varphi}\ = \ -\,\partial^2\varphi\:-\:\mu^2\varphi\:+\:\frac{\lambda}{6}\,\varphi^3\ =\ 0\;,
\end{equation}
which admits a solution with boundary conditions $\varphi|_{x_4\,\to\,\pm\,\infty}=+\,v$, $\dot{\varphi}|_{x_4\,=\,0}=0$, and $\varphi|_{|\bm{x}|\,\to\,\infty}=+\,v$, known as the ``bounce.'' In four-dimensional spherical coordinates, these boundary conditions become $\varphi|_{r\,\to\,\infty}=+\,v$ and $\dot{\varphi}|_{r\,=\,0}=0$, and the solution is the kink
\begin{equation}
\varphi \ = \ v\,\tanh\gamma (r-R)\;,
\end{equation}
where $\gamma=\mu/2$. In the language of Langer's theory of first-order phase transitions~\cite{Langer:1967ax}, the bounce corresponds to a saddle point of the free energy, which takes the system from an initially homogeneous state of false vacuum ($\varphi=+\,v$) to another that is infinitesimally close to the energetically more favorable and nominally inhomogeneous state in which a critical bubble of true vacuum ($\varphi=-\,v$) is nucleated. The radius $R$ of the bubble is then found by maximizing the free energy, or equivalently, maximizing the bounce action $B^{(0)}\equiv S[\varphi]$ with respect to this radius.

The tunneling rate is calculated by performing a saddle-point evaluation of the partition function
\begin{equation}
\label{eq:functionalint}
Z[0]\ =\ \int\![\D \Phi]\; e^{-\,S[\Phi]}\;,
\end{equation}
expanded around the bounce $\varphi$. This yields an integral over quadratic fluctuations, whose eigenspectrum is that of the fluctuation operator
\begin{equation}
G^{-1}_{xy}(\varphi)\ =\ \frac{\delta^2 S[\Phi]}{\delta \Phi_x\delta \Phi_y}\bigg|_{\Phi\,=\,\varphi}\ =\ \delta^{(4)}_{xy}\bigg[-\,\partial^2\:-\:\mu^2\:+\:\frac{\lambda}{2}\,\varphi^2\bigg]\;,
\end{equation}
where $\delta^{(4)}_{xy}\equiv\delta^{(4)}(x-y)$ is the Dirac delta function. The spectrum of this operator is, however, not positive definite. Specifically, it contains one negative eigenvalue
\begin{equation}
\label{eq:negative}
\lambda_0\ =\ \frac{1}{B^{(0)}}\frac{\delta^2 B^{(0)}}{\delta R^2}\ =\ -\,\frac{3}{R^2}\;,
\end{equation}
corresponding to dilatations of the bounce, and four zero eigenvalues, resulting from translational invariance. Consequently, the functional integral in Eq.~\eqref{eq:functionalint} is ill defined. In order to proceed, the integral over the zero modes is traded for an integral over the collective coordinates of the bounce, and the integral over the negative mode is performed by the method of steepest descent, giving rise to a non-zero imaginary part. It is this imaginary part that is related to the tunneling rate per unit volume via
\begin{equation}
\varGamma/V\ =\ 2|\mathrm{Im}\,Z[0]|/(VT)\;,
\end{equation}
where $VT$ is the four-volume of the bounce.

Now, in order to determine the tunneling rate consistently when a bounce solution arises only as a result of radiative corrections, we consider the two-particle irreducible (2PI) Cornwall-Jackiw-Tomboulis effective action~\cite{Cornwall:1974vz}, given by ($\hbar=1$)
\begin{align}
\Gamma[\bm{\psi},\bm{\Delta}]\ &=\ -\,\mathrm{ln}\,Z[\bm{J},\bm{K}]+\bm{J}^{\mathsf{T}}_x\bm{\psi}_x\nonumber\\&\qquad +\big(\bm{\psi}^{\mathsf{T}}_x\bm{K}_{xy}\bm{\psi}_y
-\bm{K}_{xy}\,\bm{\Delta}^{\mathsf{T}}_{xy}\big)\,.
\end{align}
Throughout this article, we employ the DeWitt notation in which repeated continuous indices are integrated over. The partition function is given by
\begin{equation}
\label{eq:partition}
Z[\bm{J},\bm{K}]\ =\ \int\![\mathrm{d}\bm{\Psi}]\;\exp\big[-S[\bm{\Psi}]+\bm{J}^{\mathsf{T}}_x\bm{\Psi}_x+\bm{\Psi}^{\mathsf{T}}_x\bm{K}_{xy}\bm{\Psi}\big]\;.
\end{equation}
In the above, $\bm{\Psi}$ and $\bm{\psi}$ are respectively the field multiplet and the vector of one-point functions
\begin{subequations}
\begin{gather}
\bm{\Psi}^{\mathsf{T}}\ =\ \begin{pmatrix} \Phi & \bm{X}^{\mathsf{T}} \end{pmatrix}\;,\quad \bm{X}^{\mathsf{T}} \ =\ \begin{pmatrix} X_1 & X_2 & \cdots & X_N \end{pmatrix}\;,\\
\bm{\psi}^{\mathsf{T}}\ = \ \begin{pmatrix} \phi & \bm{\chi}^{\mathsf{T}} \end{pmatrix}\;,\quad \bm{\chi}^{\mathsf{T}} \ =\ \begin{pmatrix} \chi_1 & \chi_2 & \cdots & \chi_N \end{pmatrix}\;.
\end{gather}
\end{subequations}
In addition, $\bm{J}_x$ is an $(N+1)$-dimensional vector of local sources, and $\bm{K}_{xy}$ and $\bm{\Delta}_{xy}$ are $(N + 1)\times (N+1)$ matrices of sources and two-point functions, respectively. In what follows, we will indicate the elements of $\bm{J}_x$ and $\bm{K}_{xy}$ by superscripts of the fields $\Phi$ and $X$.

We evaluate the 2PI effective action using the approach presented in Ref.~\cite{Garbrecht:2015cla}. Therein, rather than eliminating the sources $\bm{J}_x$ and $\bm{K}_{xy}$ for the one-point functions $\bm{\psi}_x$ and two-point functions $\bm{\Delta}_{xy}$, as in the standard approach, we instead express the effective action entirely in terms of the physical one-point function $\varphi_x$ of the field $\Phi_x$ (since $\braket{\chi_i}=0$) and the physical two-point functions $\bm{\mathcal{G}}_{xy}$. The physical one-point function $\varphi_x$  is the solution to the quantum equation of motion
\begin{equation}
\label{eq:deltaS}
\frac{\delta\Gamma[\bm{\psi},\bm{\Delta}]}{\delta \phi_x}\bigg|_{\varphi,\bm{\mathcal{G}}}\ =\ \frac{\delta S[\bm{\psi}]}{\delta \phi_x}\bigg|_{\substack{\phi\,=\,\varphi\\ \chi_i\,=\,0}}-J^{\Phi}_x-\!\int_y\!K^{\Phi\Phi}_{xy}\,\varphi_y\ =\ 0\;.
\end{equation}
Here, the subscript ``$\varphi,\bm{\mathcal{G}}$'' indicates that the functional derivative of the effective action is to be evaluated at the configurations $\phi=\varphi$ and $\bm{\Delta}=\bm{\mathcal{G}}$. The physical two-point functions $\bm{\mathcal{G}}_{xy}$ are the solutions to
\begin{equation}
\label{eq:SD}
\bm{\mathcal{G}}^{-1}_{xy}\ =\ \bm{G}^{-1}_{xy}\:-\:\bm{K}_{xy}\;,
\end{equation}
where
\begin{equation}
\bm{G}^{-1}_{xy}\ = \ \frac{\delta^2 S[\bm{\psi}]}{\delta \bm{\psi}^{\mathsf{T}}_x\delta \bm{\psi}_y}\bigg|_{\substack{\phi\,=\,\varphi\\ \chi_i\,=\,0}}\;.
\end{equation}

In this alternative evaluation of the effective action, the physical limit is that in which the sources are necessarily non-vanishing. This ensures that Eq.~\eqref{eq:deltaS} is self-consistent and that Eq.~\eqref{eq:SD} corresponds to the usual 2PI Schwinger-Dyson equation. Specifically, we require
\begin{subequations}
\begin{align}
J^{\Phi}_x\ &\equiv \ J^{\Phi}_{x}[\bm{\psi},\bm{\Delta}]\nonumber\\ &=\ \bigg[-\,\frac{\delta \Gamma_1[\bm{\psi},\bm{\Delta}]}{\delta \phi_x}\:+\:2\,\frac{\delta\Gamma_2[\bm{\psi},\bm{\Delta}]}{\delta\mathcal{G}^{\Phi\Phi}_{xy}}\,\phi_y\bigg]_{\varphi,\bm{\mathcal{G}}}\;,\\
K^{\Phi\Phi}_{xy}\ &\equiv\ K^{\Phi\Phi}_{xy}[\bm{\psi},\bm{\Delta}]\ =\ -\:2\,\frac{\delta\Gamma_2[\bm{\psi},\bm{\Delta}]}{\delta\Delta^{\Phi\Phi}_{xy}}\bigg|_{\varphi,\bm{\mathcal{G}}}\;,
\end{align}
\end{subequations}
where $\Gamma_1[\varphi,\bm{\mathcal{G}}]$ and $\Gamma_2[\varphi,\bm{\mathcal{G}}]$ are the one- and two-loop irreducible corrections to the effective action. As a result and in stark contrast to the standard treatment of the 2PI effective action, the saddle-point evaluation of the path integral in Eq.~\eqref{eq:partition} is performed along the \emph{quantum} trajectory of the physical one- and two-point functions $\varphi$ and $\bm{\mathcal{G}}$. Most importantly, this means that we must perform the Gaussian integral with the kernel $\bm{\mathcal{G}}^{-1}$, that is the quantum fluctuation operator. The significance of this is as follows. When false vacua emerge only as a result of radiative corrections, the classical fluctuation operator $G^{-1}$ has a positive-definite spectrum. On the other hand, the quantum fluctuation operator $\mathcal{G}^{-1}$ does not have a positive-definite spectrum, containing the expected negative and zero eigenvalues. Thus, in this alternative approach, the evaluation of the functional integral proceeds in complete analogy to the case where the SSB potential arises at tree level. The tunneling rate per unit volume is related to the imaginary part of the effective action via
\begin{equation}
\varGamma/V\ =\ 2|\mathrm{Im}\,e^{-\Gamma[\varphi,\mathcal{G}]}|/(VT)\;.
\end{equation}

\begin{figure}
\centering
\includegraphics[scale=0.45]{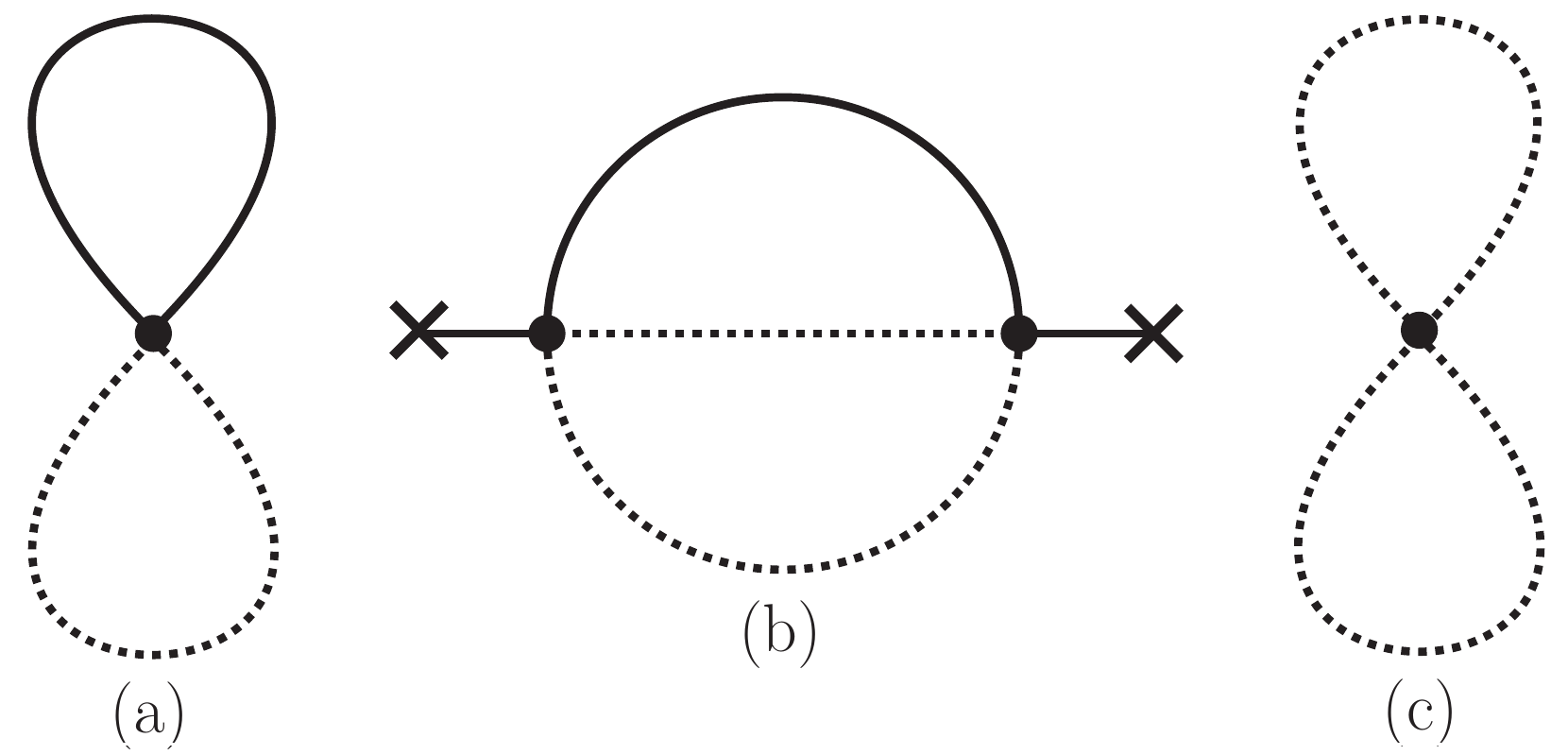}
\caption{\label{fig:Feyns}The three two-loop diagrams: (a) and (b) are $\mathcal{O}(\lambda N)$ and (c) is $\mathcal{O}(\kappa N^2)$. Solid lines correspond to $\Phi$ Green's functions and dashed lines to $X$ Green's functions; insertions of the background field $\varphi$ are marked with a cross.}
\end{figure}

At two loops, there are three diagrammatic contributions to the 2PI effective action. These are shown in Fig.~\ref{fig:Feyns}. The one-loop diagram arising from functionally differentiating Fig.~\ref{fig:Feyns}(a) with respect to the $\Phi$ Green's function appears in the equation of motion for the bounce. It is this diagram that is responsible for generating the global minimum of the potential and thereby triggering SSB. The one-loop $\Phi$ self-energies that arise from Figs.~\ref{fig:Feyns}(a) and (b) appear in the Klein-Gordon operator of the $\Phi$ field. For the positive-definite modes, these loop corrections are higher order and can safely be neglected. On the other hand, these diagrams cannot be neglected for the non-positive modes, since they provide corrections from fluctuations along directions where the effective action is either minimized or flat. As a result, by well-known arguments~\cite{Callan:1977pt}, the path integral cannot be evaluated as a Gaussian in a straightforward manner. The contribution of the zero mode is then enhanced by the (macroscopic) four-volume of the bounce $VT$, and the negative mode contributes an order-one correction to the imaginary part of the effective action. It is these diagrams that are responsible for introducing negative and zero eigenvalues to the spectrum of the $\Phi$ fluctuation operator. Thus, these loop corrections must be accounted for in the analysis of the negative semi-definite modes.  Finally, the one-loop $X$ self-energies obtained by functionally differentiating Figs.~\ref{fig:Feyns}(a)--\ref{fig:Feyns}(c) appear in the Klein-Gordon operator of the $X$ field. There, the $X$ self-energies are subleading compared to the contribution from the tree-level potential and can also safely be neglected. Hence, we may consistently consider the equation of motion for the bounce, the $X$ field Green's function and the positive-definite modes of the $\Phi$ fluctuation operator at the 1PI level, employing the 2PI approach only to get the leading behavior of the negative and zero modes. This truncation has the advantage that the $\Phi$ Green's function conveniently decouples from the problem, as we will see explicitly in what follows.

The equation of motion for the bounce takes the form
\begin{equation}
\label{eq:bounce}
-\:\partial^2\varphi\:+\:\Pi(\varphi)\varphi\ = \ 0\;,
\end{equation}
where
\begin{equation}
\label{eq:tadpole}
\Pi_x(\varphi)\ =\ \frac{\lambda N}{2}\,S_{xx}(\varphi)\;.
\end{equation}
is the one-loop tadpole diagram, which can be obtained by functionally differentiating Fig.~\ref{fig:Feyns}(a) with respect to the $\Phi$ Green's function. In four-dimensional spherical coordinates, the bounce is then the solution to
\begin{equation}
\label{eom}
-\,\frac{\D^2 \varphi}{\D r^2}-\frac{3}{r}\frac{\D\varphi}{\D r}+\Pi(\varphi)\varphi = 0\;,
\end{equation}
satisfying the boundary conditions $\varphi|_{r\,\to\,\infty}=+\,v$ and $\D\varphi/\D r|_{r\,=\,0}=0$. The $X$ Green's function is the solution to the inhomogeneous Klein-Gordon equation
\begin{equation}
\label{eq:chiprop}
\Big[-\:\partial^2\:+\:\frac{\lambda}{2}\,\varphi^2\Big]S_{xy}(\varphi)\ =\ \delta_{xy}^{(4)}\;.
\end{equation}
Finally, the effective action takes the form
\begin{align}
\label{eq:effact}
&\Gamma[\varphi,\bm{\mathcal{G}}]\ =\ B^{(0)}[\varphi]\:+\:B^{(1)}[\varphi]\:+\:\frac{i\pi}{2}\nonumber\\&\qquad-\:\frac{1}{2}\ln\bigg[|4\lambda_0|^{-1}(VT)^2(2\pi \mathcal{N}^2)^{-4}(4\gamma^2)^5\bigg]\;.
\end{align}
Here,
\begin{equation}
B^{(0)}[\varphi]\ =\ S[\Phi]\big|_{\substack{\Phi\,=\,0\\X_i\,=\,0}}\ =\ \int_x\,\bigg[\frac{1}{2}\,\big(\partial_{\mu}\varphi\big)^2\:+\:\frac{g}{3!}\big(\varphi^3-v^3\big)\bigg]\;.
\end{equation}
is the classical bounce action, and
\begin{equation}
\label{eq:B1}
B^{(1)}[\varphi]\ =\ \frac{1}{2}\,\mathrm{tr}^{(5)}\Big[\ln\,\mathrm{det}_{ij}\,\bm{\mathcal{G}}^{-1}(\varphi)\:-\:\ln\,\mathrm{det}_{ij}\,\bm{\mathcal{G}}^{-1}(v)\Big]
\end{equation}
contains the one-loop corrections. The latter have been normalized to those of the theory evaluated in the homogeneous false vacuum, so that the false vacuum has zero energy density also at the one-loop level. In Eq.~\eqref{eq:B1}, $\mathrm{det}_{ij}$ denotes the determinant in field space, and the superscript ``$(5)$'' indicates that the trace does not include the negative and zero eigenmodes of the $\Phi$ fluctuation operator. The imaginary part and the final logarithm in Eq.~\eqref{eq:effact} arise from dealing with the negative-semi-definite eigenvalues. Specifically, $\lambda_0$ is the negative eigenvalue, and the factor of $(2\pi \mathcal{N}^2)^{-4}$ results from the normalization of the zero modes. These prefactors will be discussed in Sec.~\ref{sec:bounceaction} [see Eqs.~\eqref{eq:negative2} and~\eqref{eq:N2}]. The four-volume factors $(VT)^2$ arise from integrating over the collective coordinates of the bounce, and the factor of $(4\gamma^2)^5$ is included for normalization~\cite{Konoplich:1987yd}. As already identified, the 2PI approach is needed only for the negative-semi-definite modes of the $\Phi$ fluctuation operator, in which case the one-loop correction simplifies to
\begin{equation}
B^{(1)}[\varphi]\ =\ \frac{N}{2}\,\mathrm{tr}\Big[\ln\,S^{-1}(\varphi)\:-\:\ \ln\,S^{-1}(v)\Big]\;.
\end{equation}

Thus, our goal is straightforward: calculate the self-consistent form of the 1PI $X$ field Green's function in the background of the quantum bounce $\varphi$ via the coupled system in Eqs.~\eqref{eq:bounce} and~\eqref{eq:chiprop}. We remark again that the $\Phi$ Green's function conveniently decouples from the problem.

\section{Numerical procedure}
\label{sec:numproc}

In order to simplify matters, we employ the approximation scheme outlined in~Ref.~\cite{Garbrecht:2015oea}:
\begin{itemize}

\item Thin wall---when the minima are quasi-degenerate (i.e.~the cubic coupling $g$ is chosen to be sufficiently small) and, as a result, the radius of the bubble $R$ is much larger than the width of the bubble wall, we may neglect the damping terms [$(-3/r) \,\mathrm{d}/\mathrm{d}r$] in Eqs.~\eqref{eom} and \eqref{eq:chiprop}.

\item Planar wall---when $R$ is large, we may approximate the sum over discrete angular momenta $j(j+2)\hbar$ by an integral over a continuous momentum $\mathbf{k}$. In so doing, we replace the hyperspherical coordinates $(r,\theta,\phi,\psi)$ by the Cartesian coordinates $(z,\bm{z}_{\parallel})$, where $z$ is oriented normal to the bubble wall and the three-vector $\bm{z}_{\parallel}$ lies within the hyperplane of the bubble wall.

\end{itemize}

Next, we make the change of variables
\begin{equation}
u\ =\ \tanh\,\gamma z\;,
\end{equation}
in order to map the infinite domain of the variable $z$ to the finite interval $[-1,+1]$.
The mass parameter $\gamma > 0$ is defined from the second derivative of the Coleman-Weinberg effective potential, evaluated in the homogeneous false vacuum. Specifically,
\begin{equation}
\label{eq:gamma2}
\gamma^2\ =\ U_{\mathrm{eff}}^R{}''(v)/4\ =\ \frac{\lambda^2 N}{128\pi^2}\,v^2 \;,
\end{equation}
cf.~the equivalent definition of $\gamma$ in the case of the $\lambda\Phi^4$ theory with tachyonic mass~\cite{Garbrecht:2015oea}.

If the profile of the bubble wall were a pure hyperbolic tangent, $u$ would in fact be the normalized bounce $\varphi/v$. As we will see, for the model under consideration, the profile of the bubble wall differs only marginally from this form. Thus, for a fixed scale $M$ and ratio of couplings $\rho$, the gradients of the bounce in the vicinity of the bubble wall will scale like $\lambda\sqrt{N}$ for $N\gg 2$.

\subsection{Iterative procedure}

In order to find the self-consistent solution for the bounce and $X$ field Green's function from Eqs.~\eqref{eq:bounce} and \eqref{eq:chiprop}, we employ an iterative procedure. This proceeds as follows:
\begin{enumerate}

\item We calculate a first approximation to the bounce by promoting the homogeneous field configuration, appearing in the Coleman-Weinberg effective potential, to a spacetime-dependent configuration. Hence, in the physical coordinates $(z,\mathbf{z}_{\parallel})$, the equation of motion for the bounce in this first iteration takes the form
\begin{equation}
-\,\partial_z^2\varphi_z\:+\:U^R_{\mathrm{eff}}{}'(\varphi_z)\ =\ 0\;.
\end{equation} 

\item We then insert the solution for the bounce into Eq.~\eqref{eq:chiprop} and solve for the $X$ field Green's function.

\item Next, we take the coincident part of the $X$ Green's function to calculate the tadpole correction in Eq.~\eqref{eq:tadpole}, renormalizing in the homogeneous false vacuum.

\item The tadpole correction can now be inserted into Eq.~\eqref{eq:bounce}, and we solve again for the bounce, iterating over steps $2$--$4$ until the results have converged.

\end{enumerate}

\subsection{Bounce action}
\label{sec:bounceaction}

In order to determine the bounce action and the negative eigenvalue, we need first to find the bubble radius $R$. This is done by minimizing the energy difference between the latent heat of the bubble and its surface tension. Isolating these contributions, we may write the full bounce action in the form
\begin{equation}
B\ =\ B^{(0)}\:+\:B^{(1)}\ =\ B_{\mathrm{surface}}\:+\:B_{\mathrm{vacuum}}\;.
\end{equation}

The surface tension scales like $R^3$ and arises from the kinetic term and fluctuation determinant:
\begin{equation}
B_{\mathrm{surface}}\ =\ 2\pi^2R^3\bigg[\int_{R-\epsilon}^{R+\epsilon}\D z\;\frac{1}{2}\,\bigg(\frac{\D \varphi}{\D z}\bigg)^2\:+\:\overline{B}^{(1)}\bigg]\;,
\end{equation}
where
\begin{equation}
\overline{B}^{(1)}\ =\ \frac{B^{(1)}}{2\pi^2R^3}\;.
\end{equation}

On the other hand, the latent heat scales like $R^4$. It arises solely from the $\mathbb{Z}_2$-breaking term and is given by
\begin{equation}
B_{\mathrm{vacuum}}\ =\ 2\pi^2\int_0^{R-\epsilon}\D r\;r^3 U(-v)\;.
\end{equation}
Since $U(-v) = -\,gv^3/3$, we obtain the analytic result
\begin{equation}
B_{\mathrm{vacuum}}\ =\ -\:\frac{\pi^2R^4gv^3}{6}\;.
\end{equation}

Extremizing the action with respect to $R$, we find
\begin{equation}
\label{R}
R\ =\ \frac{9}{gv^3}\bigg[\int_{R-\epsilon}^{R+\epsilon}\D z\;\frac{1}{2}\,\bigg(\frac{\D \varphi}{\D z}\bigg)^2\:+\:\overline{B}^{(1)}\bigg]\;.
\end{equation}
Note that we have neglected contributions arising from the variation of $\varphi$ and $\overline{B}^{(1)}$ with respect to $R$, which are, in the thin-wall approximation, subleading compared to the variations of the area and volume factors.

The integral over the kinetic term can be evaluated numerically by writing
\begin{align}
& \int_{R-\epsilon}^{R+\epsilon}\D z\;\frac{1}{2}\,\bigg(\frac{\D \varphi}{\D z}\bigg)^2\ =\ \frac{1}{2}\int_{-v}^{+v}\D\varphi\;\frac{\D \varphi}{\D z}\nonumber\\ & \qquad = \ \frac{\gamma}{2} \int_{-1}^{+1}\D u\;(1-u^2)\bigg(\frac{\D \varphi}{\D u}\bigg)^2\;.
\end{align}
In order to obtain $\overline{B}^{(1)}$, we employ the method due to Baacke and Junker~\cite{Baacke:1993jr,Baacke:1993aj,Baacke:1994ix} (see also Refs.~\cite{Baacke:1994bk,Baacke:2008zx}) for calculating the fluctuation determinant from direct integration of the Green's function (see Appendix~\ref{app:Baacke}). By this means, we may express
\begin{align}
\overline{B}^{(1)}[\varphi]\ &=\ -\:\frac{N}{2}\int_{-1}^{+1}\frac{\mathrm{d} u}{\gamma (1-u^2)}\nonumber\\&\quad \times \int_{0}^{\Lambda}\mathrm{d}k\int_{0}^{\Lambda^2}\mathrm{d}s\;\frac{k^2}{2\pi^2}\,\widetilde{S}(u,k^2+s;\varphi)\;,
\end{align}
where $s\in\mathbb{R}$ is an auxiliary parameter and we have defined the normalized Green's function
\begin{equation}
\widetilde{S}(u,k^2+s;\varphi)\ =\ S(u,k^2+s;\varphi)\:-\:S(1,k^2+s;\varphi)\;.
\end{equation}
The dependence on the UV cutoff $\Lambda$ is removed by the addition of the normalized counterterm
\begin{equation}
\label{B1counter}
\delta B^{(1)}\ =\ \int\!\mathrm{d}^4x\bigg[\frac{1}{2!}\,\delta m_{\varphi}^2\big(\varphi^2-v^2\big)\:+\:\frac{1}{4!}\,\delta\alpha\,\big(\varphi^4-v^4\big)\bigg]\;,
\end{equation}
where the mass and coupling counterterms $\delta m_{\varphi}^2$ and $\delta\alpha$ are given in Appendix~\ref{app:CW}.

Substituting Eq.~\eqref{R} back into the expression for the bounce action, we may show that
\begin{equation}
B\ =\ \frac{\pi^2gv^3}{18}\,R^4\;.
\end{equation}
In addition, the negative eigenvalue is given by
\begin{equation}
\label{eq:negative2}
\lambda_0\ =\ -\,\frac{3}{R^2}\;,
\end{equation}
in complete analogy to the tree-level case. The latter fact can readily be verified by differentiating the equation of motion for the bounce with respect to $r$, yielding
\begin{align}
&\bigg[-\,\frac{\D^2}{\D r^2}\:-\:\frac{3}{r}\,\frac{\D}{\D r}\:+\:\Pi_r\bigg]\frac{\D \varphi}{\D r}\nonumber\\&\qquad +\:\int\!\D^4 x'\; \varphi_r\,\frac{\delta \Pi_r}{\delta\varphi_{r'}}\,\frac{\D \varphi_{r'}}{\D r'}\ =\ -\:\frac{3}{r^2}\,\frac{\D \varphi_r}{\D r}\;.
\end{align}
Note that the term arising from varying the tadpole diagram is non-local, resulting in an additional convolution integral. In the thin-wall approximation, we may set $r=R$ in those terms originating from the damping term, giving the eigenequation for the negative mode $\phi_0=\mathcal{N}\partial_r\varphi$ with the eigenvalue given in Eq.~\eqref{eq:negative2}.

Finally, the normalization $\mathcal{N}$ of the zero modes $\phi_{\mu}=\mathcal{N}\partial_{\mu}\varphi$ is given by
\begin{equation}
\label{eq:N2}
\mathcal{N}^{-2}\ =\ \frac{1}{4}\int_x\bigg(\frac{\D \varphi}{\D z}\bigg)^2\ =\ \frac{1}{2}\,\pi^2R^3\gamma\int_{-1}^{+1}\!\D u (1-u^2)\,\bigg(\frac{\D \varphi}{\D u}\bigg)^2\;,
\end{equation}
which follows from the orthonormality condition
\begin{equation}
\mathcal{N}^2\int_x\phi^*_{\mu}\phi_{\nu}\ =\ \delta_{\mu\nu}\;. 
\end{equation}

\section{Numerical results}
\label{sec:numres}

In this section, we present the numerical results of the iterative procedure outlined in Sec.~\ref{sec:numproc} for the model described in Sec.~\ref{sec:model}.

In order to provide independent cross-checks of the numerical results, the iterative procedure was performed using two distinct approaches: the first employed the built-in differential solvers of Mathematica and the second was based upon Chebyshev pseudospectral collocation methods (see e.g.~Ref.~\cite{Boyd}). In the latter, the equation for the bounce was linearized using the Newton-Kantarovich method (see e.g.~Ref.~\cite{Boyd}).

The renormalization of the tadpole correction and the one-loop correction $B^{(1)}$ was performed by constructing momentum-dependent pseudo-counterterms from the analytic expressions in Appendix~\ref{app:CW}. These were subtracted at the level of the integrands, thereby avoiding residual cutoff dependent terms resulting from errors in the numerical integration.

The numerical analysis was repeated for a range of $\lambda$ and $N$ consistent with $0.04\leq\lambda^2N\leq0.4$. The upper limit was imposed so as to remain within the perturbative regime of the large $N$ expansion~\cite{'tHooft:1973jz}. This limit was identified numerically by comparing the relative contributions of the $X$ and $\Phi$ fluctuation determinants. Parameter points consistent with these limits were chosen from the sets $\lambda=\{0.03,0.04,0.05,0.06,0.07,0.08,0.09,0.10\}$ and $N=\{4,8,12,16,20,24,28,32,36,40\}$, giving a total of $55$ sample points. A fixed ratio $\rho=3$ and mass scale $M=1$ were used throughout. The numerical results converged sufficiently after two iterations, amounting to including the first back-reaction of the gradient effects on the bounce configuration itself.

\begin{figure}
\centering
\includegraphics[scale=0.6]{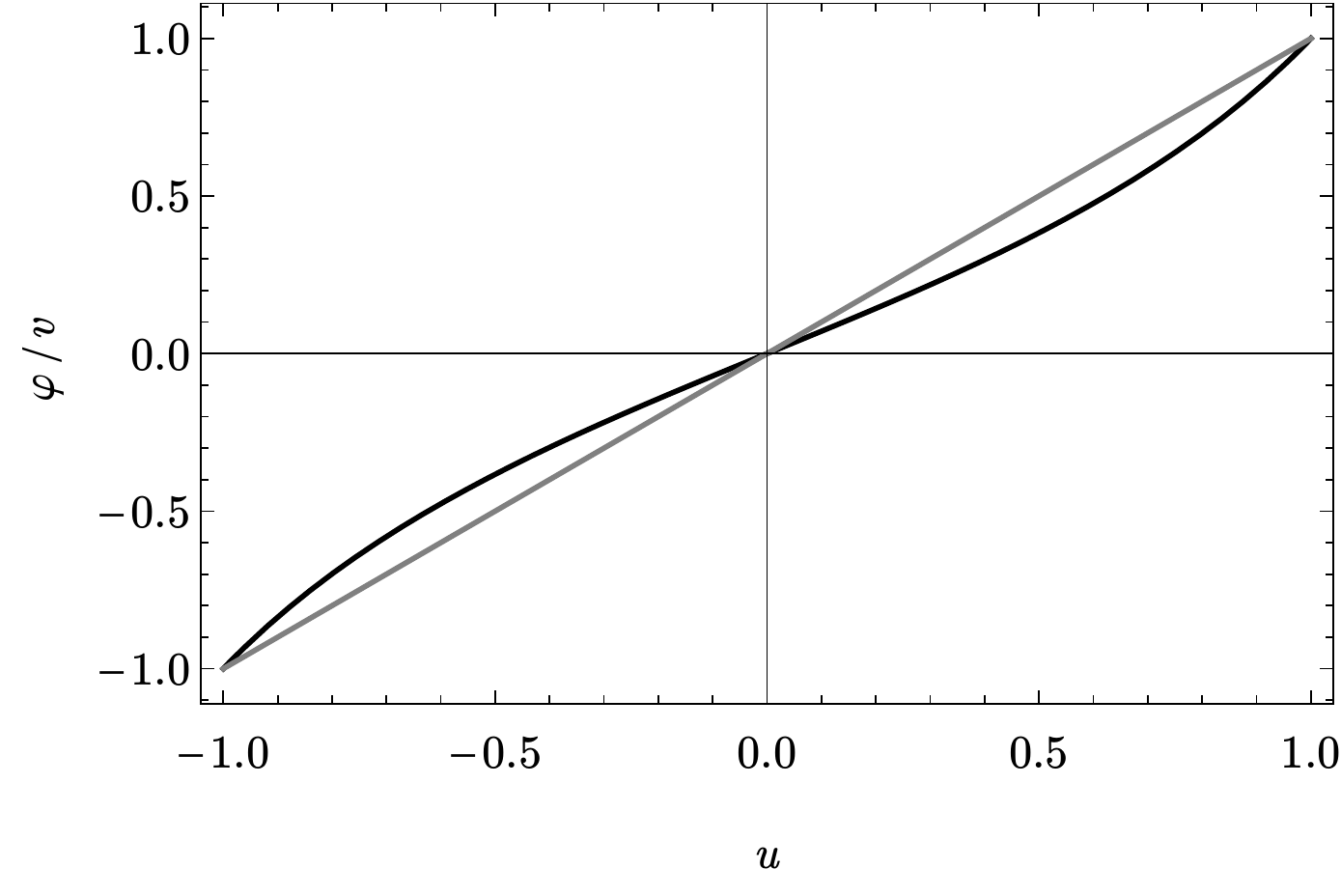}
\caption{\label{fig:bounce}Plot of the normalized bounce $\varphi/v$ as a function of $u$ for $\lambda=0.1$ and $N=40$. The straight line corresponds to a pure hyperbolic tangent profile.}
\end{figure}

In Fig.~\ref{fig:bounce}, we plot the profile of the bounce as a function of the transformed coordinate $u$ for the largest of the parameter choices ($\lambda=0.1$ and $N=40$). Therein, we see the marginal departure of the bounce from a pure hyperbolic tangent.

\begin{figure}
\includegraphics[scale=0.6]{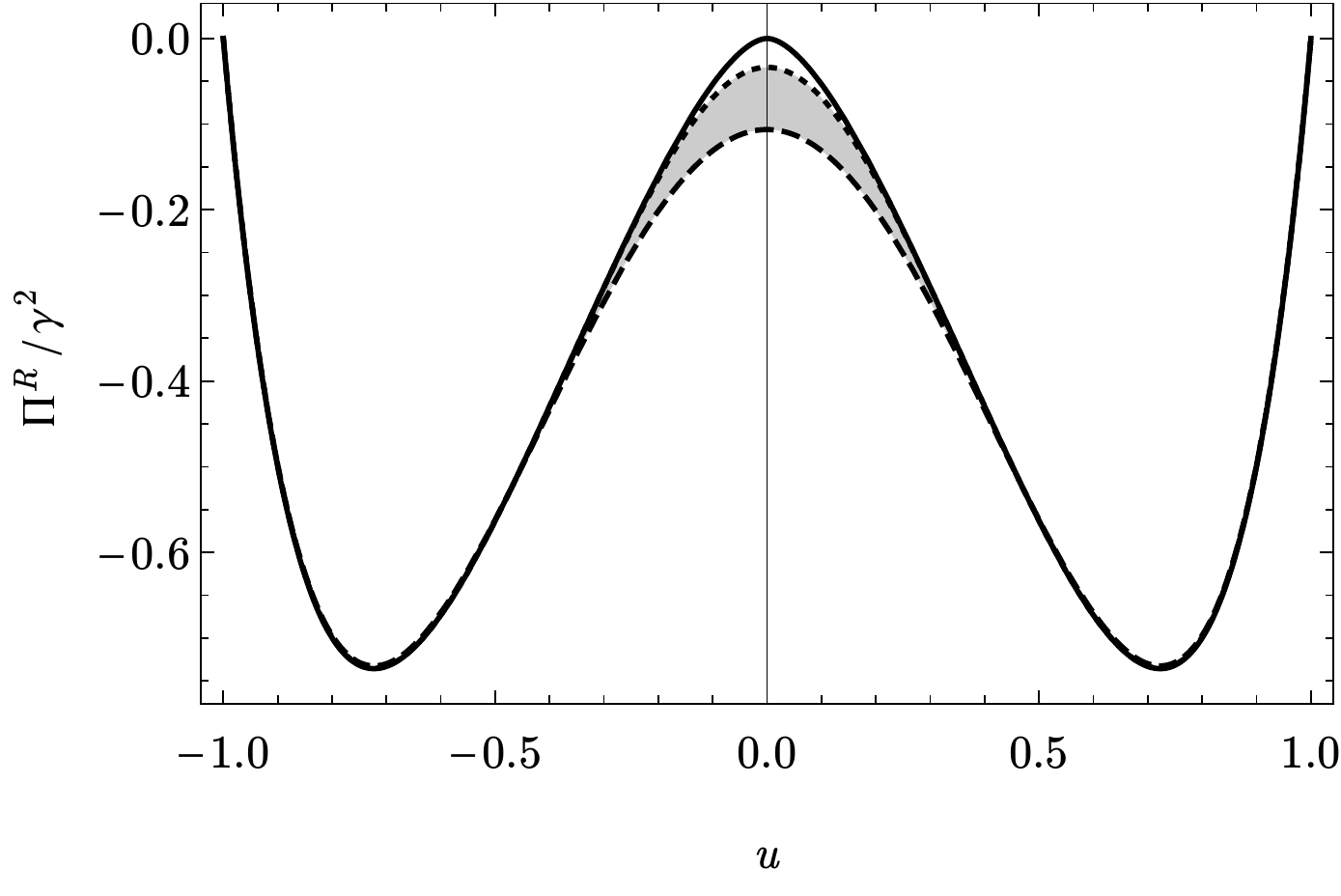}
\caption{\label{fig:piR}Plot of the renormalized tadpole $\Pi^R$, normalized to the value of $\gamma^2$, as a function of $u$. The dashed and dotted lines correspond respectively to the largest ($\lambda=0.1$, $N=40$) and smallest ($\lambda=0.1$, $N=4$) values of $\lambda^2N$ in the analyzed parameter range. The solid line is obtained from the Coleman-Weinberg effective potential and corresponds to $\mathrm{d}U_{\mathrm{eff}}/\mathrm{d}\varphi/\varphi/\gamma^2$ evaluated for the bounce of the first iteration. The shaded area indicates the variation of $\Pi^R$ over the analyzed parameter range that results from including the corrections from the gradients of the bounce.}
\end{figure}

\begin{figure}
\centering
\includegraphics[scale=0.6]{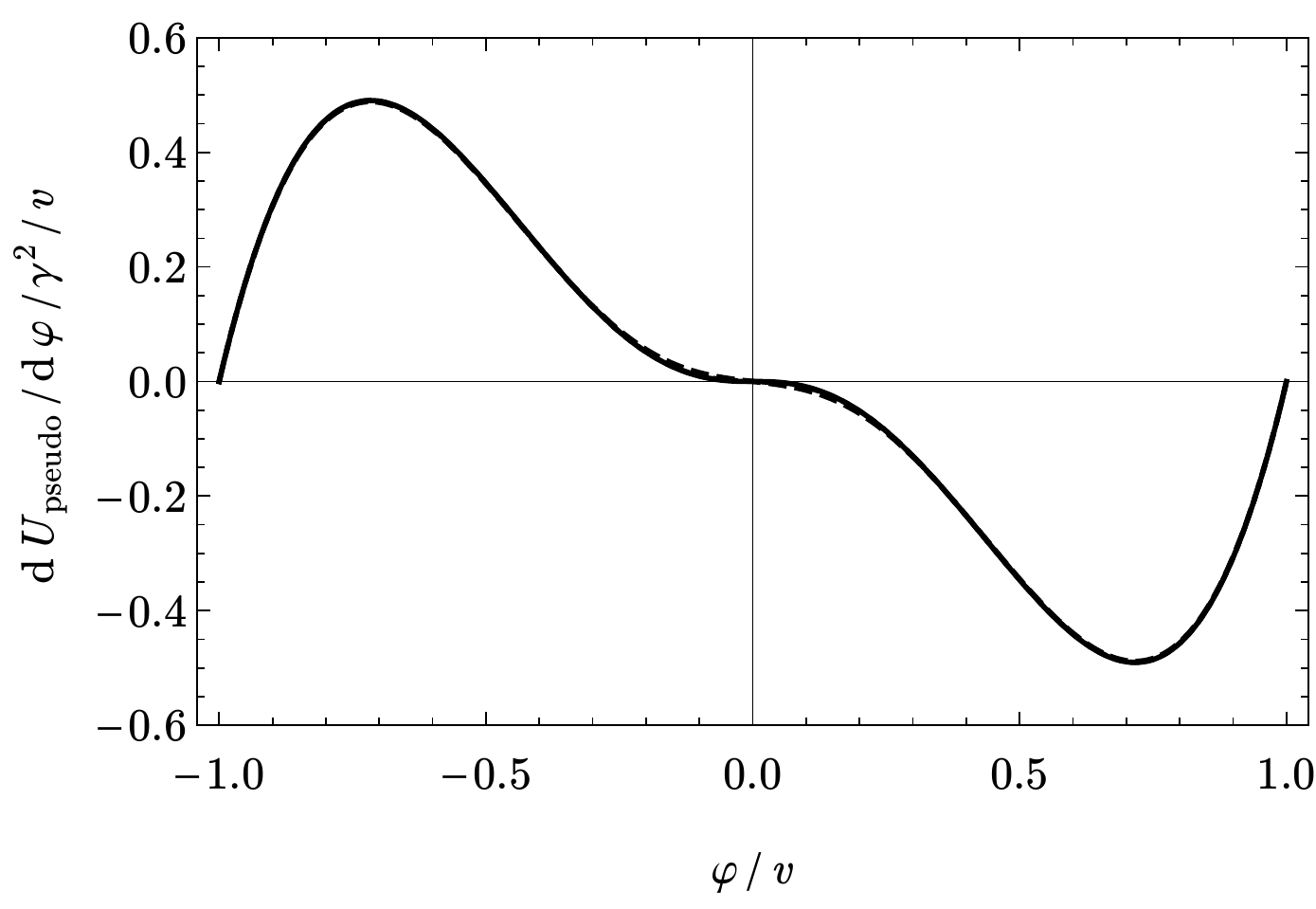}\\
(a)\vspace{1em}\\
\includegraphics[scale=0.6]{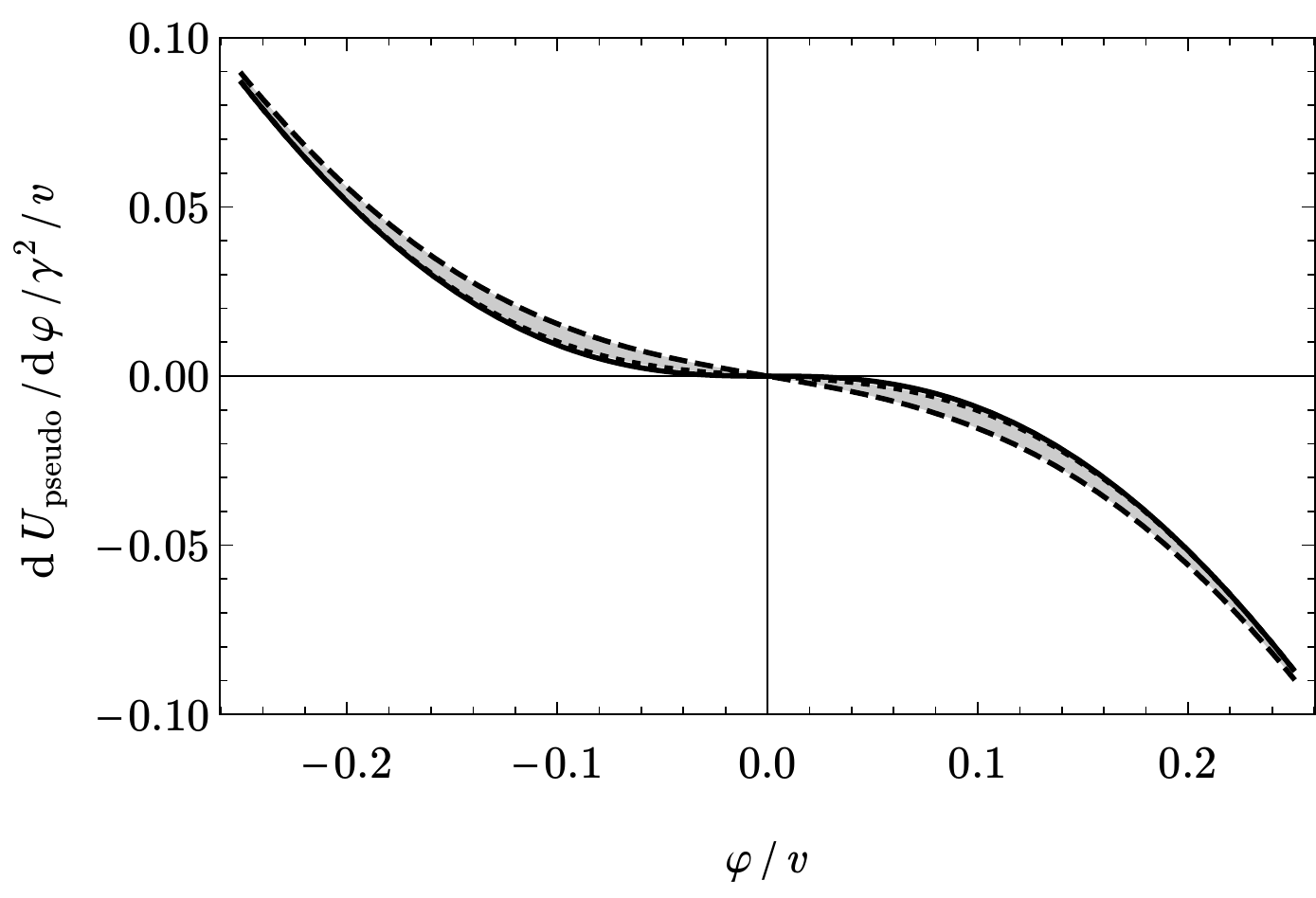}\\
(b)
\caption{\label{fig:pseudo}The first derivative of the pseudo-potential $\mathrm{d}U_{\mathrm{pseudo}}/\mathrm{d}\varphi = \varphi\,\Pi^R$, normalized to $\gamma^2v$, as a function of $u$ over (a) the full domain and (b) in the vicinity of the bubble wall. The dashed and dotted lines correspond respectively to the largest ($\lambda=0.1$, $N=40$) and smallest ($\lambda=0.1$, $N=4$) values of $\lambda^2N$ in the analyzed parameter range. The solid line is obtained from the Coleman-Weinberg effective potential and corresponds to $\mathrm{d}U_{\mathrm{eff}}/\mathrm{d}\varphi/\gamma^2/v$. The shaded area indicates the variation of $\varphi\,\Pi^R$ over the analyzed parameter range that results from including the corrections from the gradients of the bounce.}
\end{figure}

\begin{figure}[t]
\includegraphics[scale=0.6]{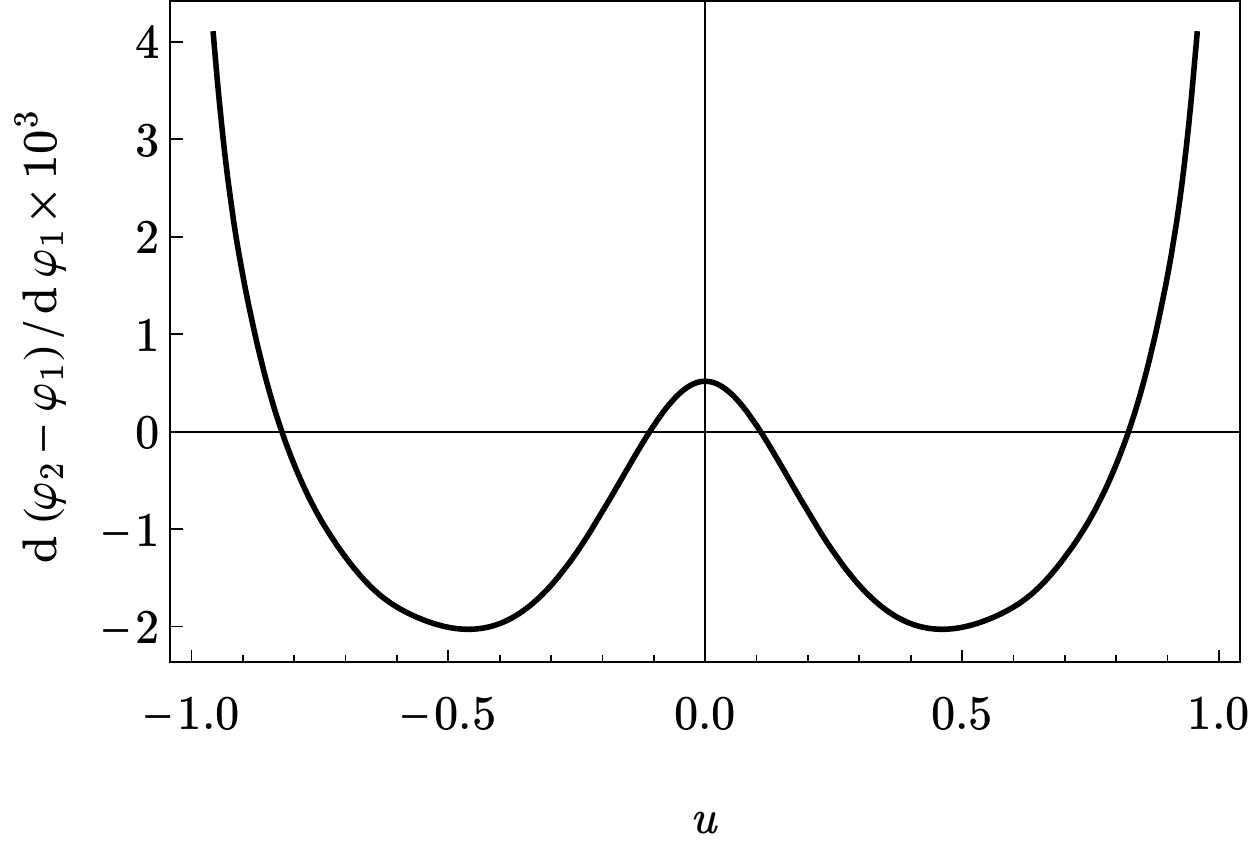}
\caption{\label{fig:grads}Plot of the difference between the gradient of the bounce from the second ($\varphi_2$) and first ($\varphi_1$) iterations divided by the gradient of the first iteration for the largest of the parameter values $\lambda=0.1$ and $N=40$.}
\end{figure}

In order to illustrate the relative importance of the gradients in the vicinity of the bubble wall, it is convenient to consider the quantities
\begin{equation}
\label{eq:pseudo1}
\frac{1}{\gamma^2}\,\frac{1}{\varphi}\,\frac{\mathrm{d}U_{\mathrm{eff}}^R}{\mathrm{d}\varphi}\ \approx\ \frac{1}{\gamma^2}\,\Pi^{R}\;,
\end{equation}
and
\begin{equation}
\label{eq:pseudo2}
\frac{1}{\gamma^2v}\,\frac{\mathrm{d}U_{\mathrm{eff}}^R}{\mathrm{d}\varphi}\ \approx\ \frac{1}{\gamma^2v}\,\varphi\,\Pi^{R}\ \equiv\ \frac{1}{\gamma^2v}\,\frac{\mathrm{d}U_{\mathrm{pseudo}}}{\mathrm{d}\varphi}\;,
\end{equation}
where the approximation results from the fact the right-hand sides include gradient effects. Equation~\eqref{eq:pseudo2} defines the pseudo-potential, which appears in the equation of motion for the bounce [Eq.~\eqref{eq:bounce}]. By comparing the left-hand sides of Eqs.~\eqref{eq:pseudo1} and~\eqref{eq:pseudo2} with the definition of $\gamma$ in Eq.~\eqref{eq:gamma2} and the renormalized effective potential in Eq.~\eqref{eq:Ueffrho}, we may verify that the left-hand sides are, for $N\gg 2$, independent of $\lambda$ and $N$. Hence, any variation seen in the plots of the right-hand sides of Eqs.~\eqref{eq:pseudo1} and~\eqref{eq:pseudo2} across the analyzed parameter range will result solely from the impact of the gradients in the vicinity of the wall. This can be seen clearly in Figs.~\ref{fig:piR} and~\ref{fig:pseudo}. Therein, the shaded regions indicate the variation in the vicinity of the bubble wall over the range of $\lambda^2 N$ compared to the homogeneous Coleman-Weinberg result, indicated by the solid lines. As is clear from Fig.~\ref{fig:pseudo}, in spite of the order-$10\,\%$ effect on the renormalized tadpole in Fig.~\ref{fig:piR}, the impact of the back-reaction of the gradients on the equation of motion for the bounce is negligible. This may be understood as follows: in the thin-wall regime, the gradients are relevant only in the vicinity of the bubble wall. In this region, however, the bounce configuration itself is going to zero, and the additional occurrence of the bounce $\varphi$ in Eq.~\eqref{eq:pseudo2} compared to Eq.~\eqref{eq:pseudo1} leads to the suppression of the gradient effects between Figs.~\ref{fig:piR} and~\ref{fig:pseudo}. As a result, over the range of parameters, negligible variation was seen in the bounce between each iteration. This is illustrated further in Fig.~\ref{fig:grads}, where we plot the difference in the gradient of the bounce between the first and second iterations divided by that of the first iteration for the largest parameter values $\lambda=0.1$ and $N=40$. We see that the back-reaction of the gradient effects leads to a correction of order $0.1$--$0.2\,\%$ in the region of interest ($u\sim\pm\,0.5$, cf.~Fig.~\ref{fig:grads}). One may conclude that, in this case, the bounce as determined in the Coleman-Weinberg effective potential gives a good approximation to the self-consistent solution.

\begin{figure}[t]
\centering
\includegraphics[scale=0.6]{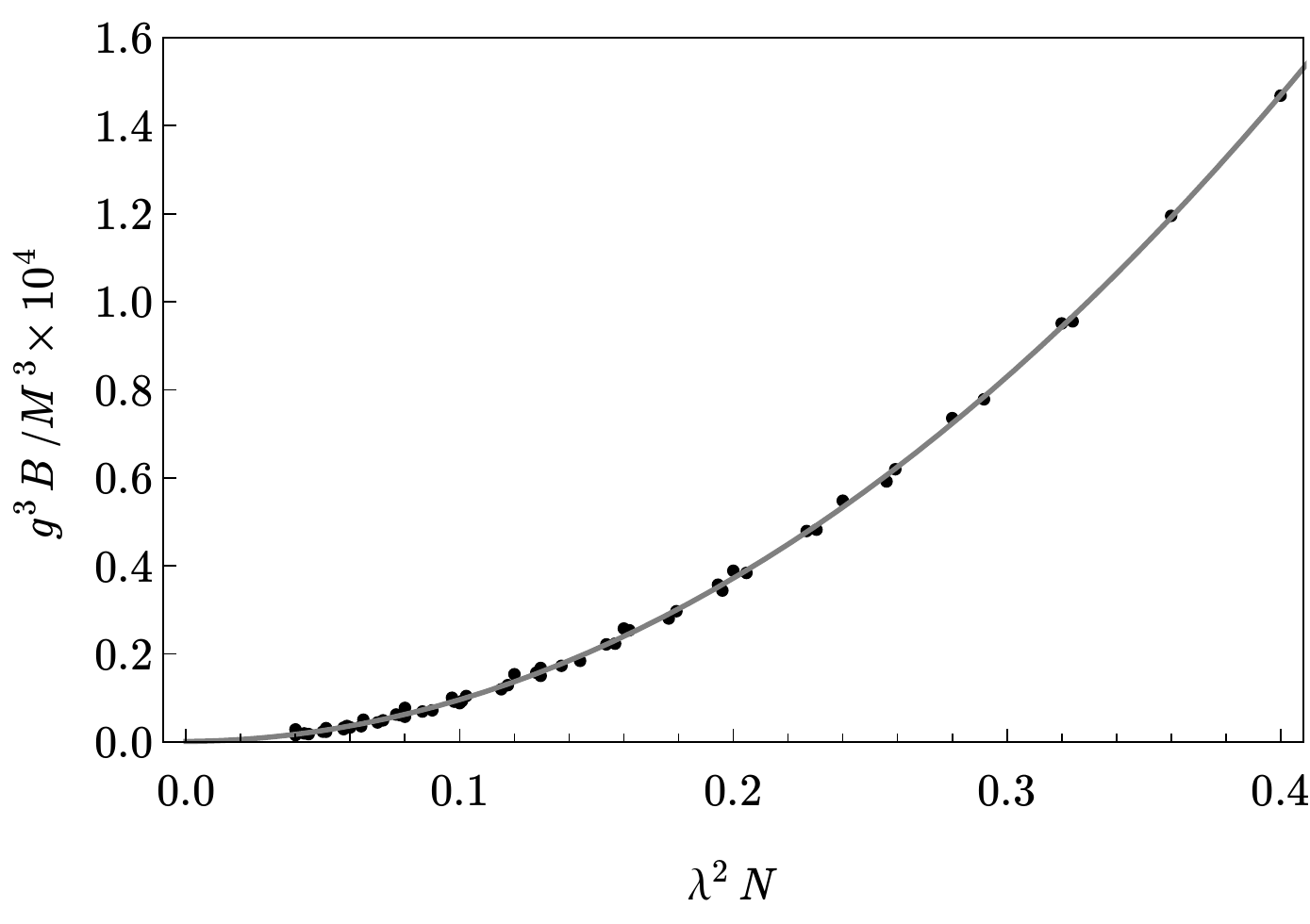}
\caption{\label{fig:B}Plot of the bounce action $B$ as a function of $\lambda^2N$. The fit corresponds to a third-order polynomial in $\lambda^2N$. The residual deviation from a polynomial in $\lambda^2N$ is anticipated to be in part a consequence of the $e^{F/(2N)}$ dependence of the vacuum expectation value $v$ [cf.~Eq.~\eqref{eq:vev} and Fig.~\ref{fig:B1comp}].}
\end{figure}

\begin{figure}[t]
\centering
\includegraphics[scale=0.6]{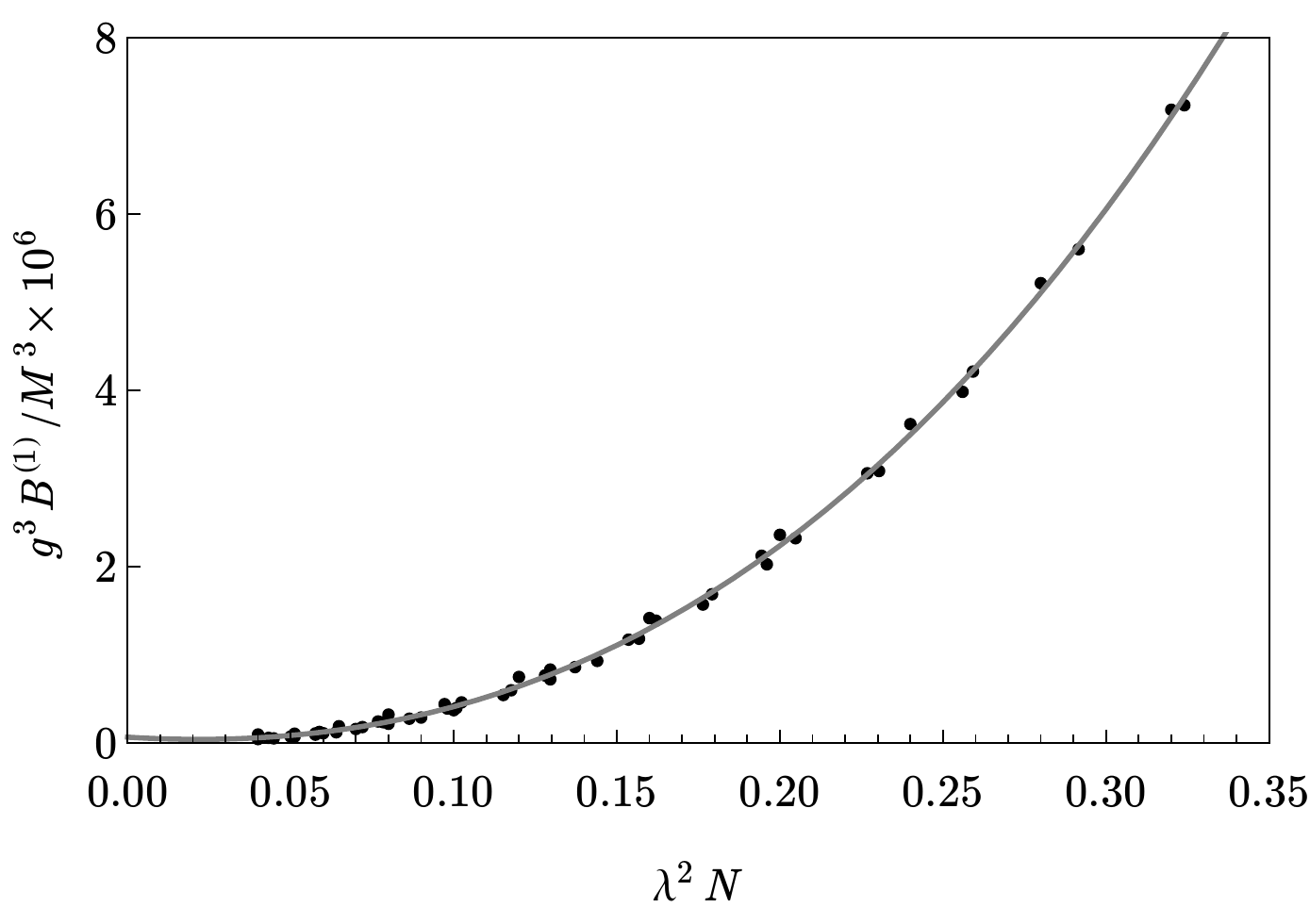}
\caption{\label{fig:B1}Plot of the one-loop correction to the bounce action $B^{(1)}$ as a function of $\lambda^2N$. The fit corresponds to a third-order polynomial in $\lambda^2N$. The residual deviation from a polynomial in $\lambda^2N$ is anticipated to be in part a consequence of the $e^{F/(2N)}$ dependence of the vacuum expectation value $v$ [cf.~Eq.~\eqref{eq:vev} and Fig.~\ref{fig:B1comp}].}
\end{figure}

\begin{figure}[t]
\centering
\includegraphics[scale=0.6]{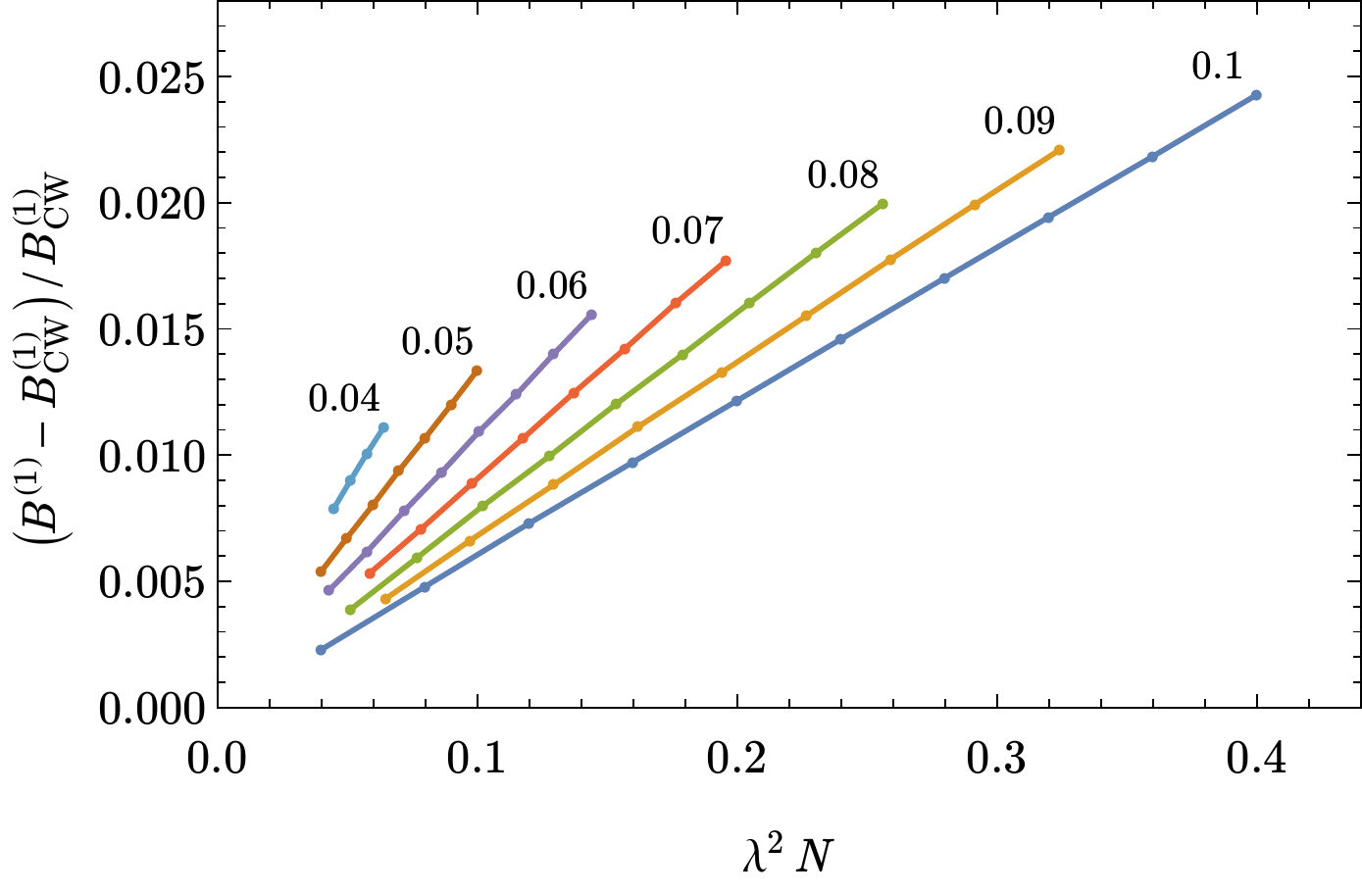}
\caption{\label{fig:B1comp}Plot of the ratio of the difference $B^{(1)}-B^{(1)}_{\mathrm{CW}}$ to $B^{(1)}_{\mathrm{CW}}$ as a function of $\lambda^2N$. The labels indicate the values of $\lambda$. The residual deviation from a polynomial in $\lambda^2N$ is anticipated to be in part a consequence of the $e^{F/(2N)}$ dependence of the vacuum expectation value $v$ [cf.~Eq.~\eqref{eq:vev}].}
\end{figure}

In Figs.~\ref{fig:B} and~\ref{fig:B1}, we plot the full bounce action $B$ and the contribution from the one-loop corrections $B^{(1)}$, multiplied by $g^3$, as a function of $\lambda^2N$. In order to illustrate the relative importance of accounting for the gradients, Fig.~\ref{fig:B1comp} shows the difference between the self-consistent one-loop contribution $B^{(1)}$ and the equivalent result calculated assuming a homogeneous background field configuration, denoted $B^{(1)}_{\mathrm{CW}}$, as a fraction of the homogeneous result. It is in the one-loop corrections that the inclusion of the gradients had the dominant absolute impact. Even so, the relative importance of the gradients is at the percent level. Hence, we see again that the one-loop Coleman-Weinberg homogeneous result provides a good approximation for the bounce action. However, in Fig.~\ref{fig:B1comp}, we observe that the difference between the inhomogeneous self-consistent and homogeneous Coleman-Weinberg fluctuation determinants does not scale as a polynomial in the expansion parameter $\lambda^2N$. In particular, there is a residual dependence on both $\lambda$ and $N$, the latter of which is approximately linear over the analyzed parameter range. This is anticipated to be in part a consequence of the $e^{F/(2N)}$ dependence of the vacuum expectation value $v$ [cf.~Eq.~\eqref{eq:vev}]. Fig.~\ref{fig:B1comp} suggests that gradient effects contribute a term to the fluctuation determinant that scales approximately like $\lambda^2N^2$. Since the two-loop self-energies scale as either $\lambda\kappa N^2$ or $\lambda^2N$, there is therefore the possibility that gradient effects may compete with two-loop effects.

\begin{figure}[t]
\centering
\includegraphics[scale=0.6]{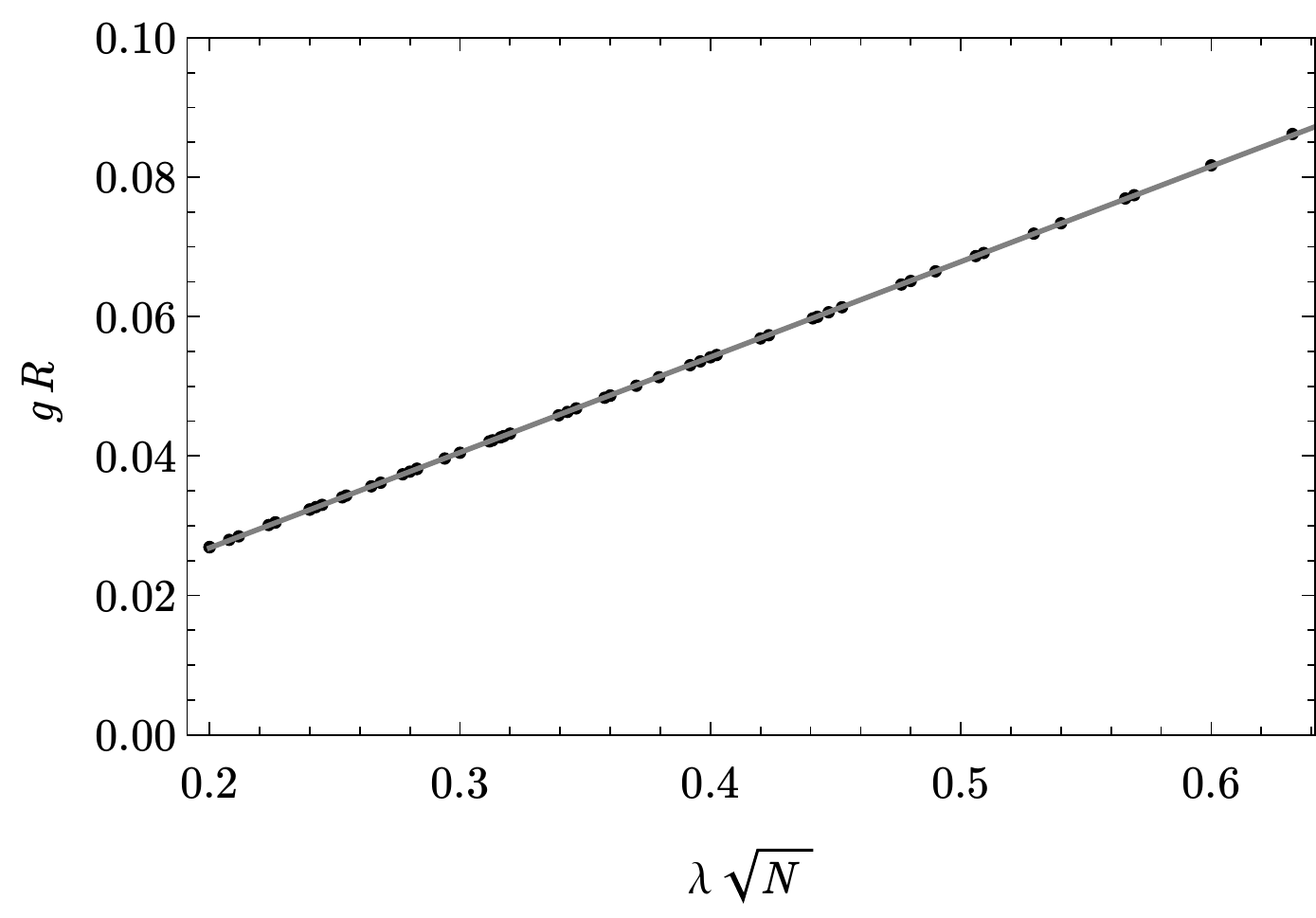}
\caption{\label{fig:R}Plot of the bubble radius $R$ times $g$ as a function of $\lambda\sqrt{N}$.}
\end{figure}

\begin{figure}[t]
\centering
\includegraphics[scale=0.6]{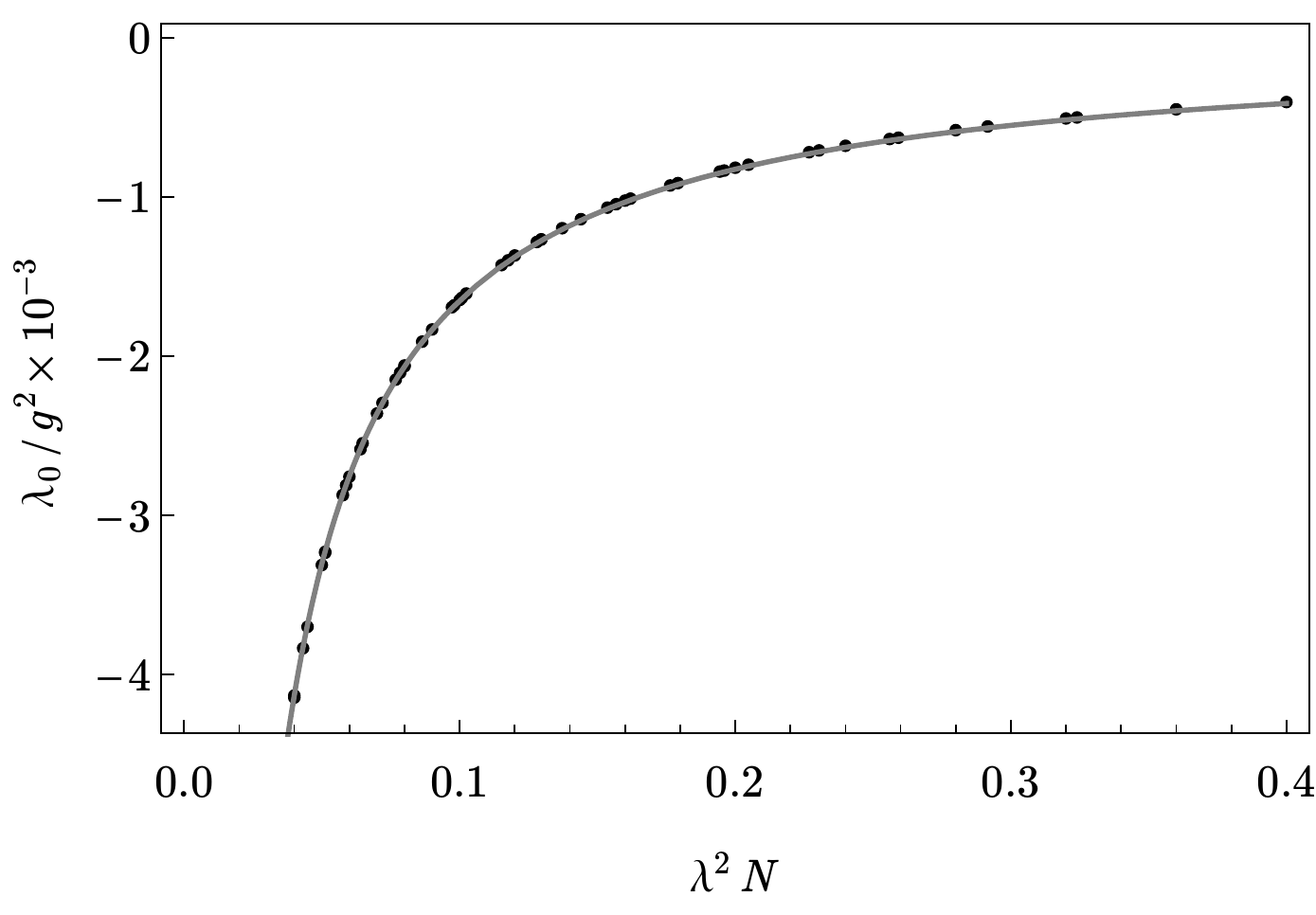}
\caption{\label{fig:lambda0}Plot of the negative eigenvalue $\lambda_0=-3/R^2$, normalized to $g^2$, as a function of $\lambda^2N$. The fit is of the form $a/\lambda^2/N$, where $a$ is a real constant.}
\end{figure}

\begin{figure}[t]
\centering
\includegraphics[scale=0.6]{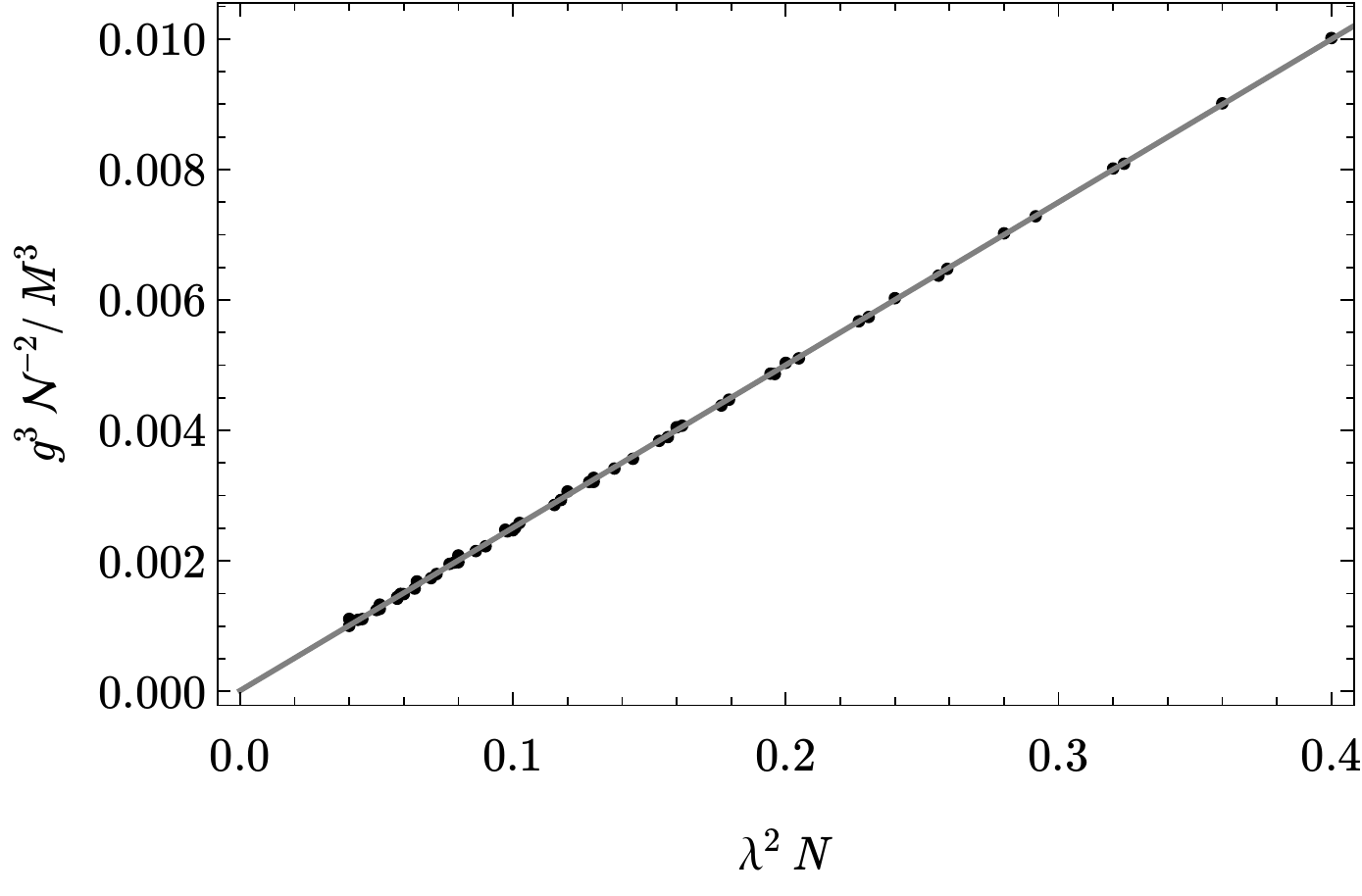}
\caption{\label{fig:N2}Plot of the squared inverse normalization of the zero modes, normalized to $g^{-3}$, as a function of $\lambda^2N$. Comparing with Eq.~\eqref{eq:N2}, we see that the gradients of the bounce scale like $\lambda\sqrt{N}$, as anticipated from Eq.~\eqref{eq:gamma2}.}
\end{figure}

In Figs.~\ref{fig:R} and~\ref{fig:lambda0}, we plot the bubble radius $R$ multiplied by $g$ and the negative eigenvalue $\lambda_0$ normalized to $g^2$. The latter is, of course, not independent and is included for completeness. In addition, in Fig.~\ref{fig:N2}, we plot the squared inverse normalization of the zero modes $\mathcal{N}^{-2}$ multiplied by $g^3$. By comparing these plots with the definition of the normalization in Eq.~\eqref{eq:N2}, we infer that the gradient with respect to $u$ scales like $\lambda\sqrt{N}$, as anticipated below Eq.~\eqref{eq:gamma2}, and that the bubble radius $R$ scales linearly with this gradient.

\begin{figure}[t]
\centering
\includegraphics[scale=0.6]{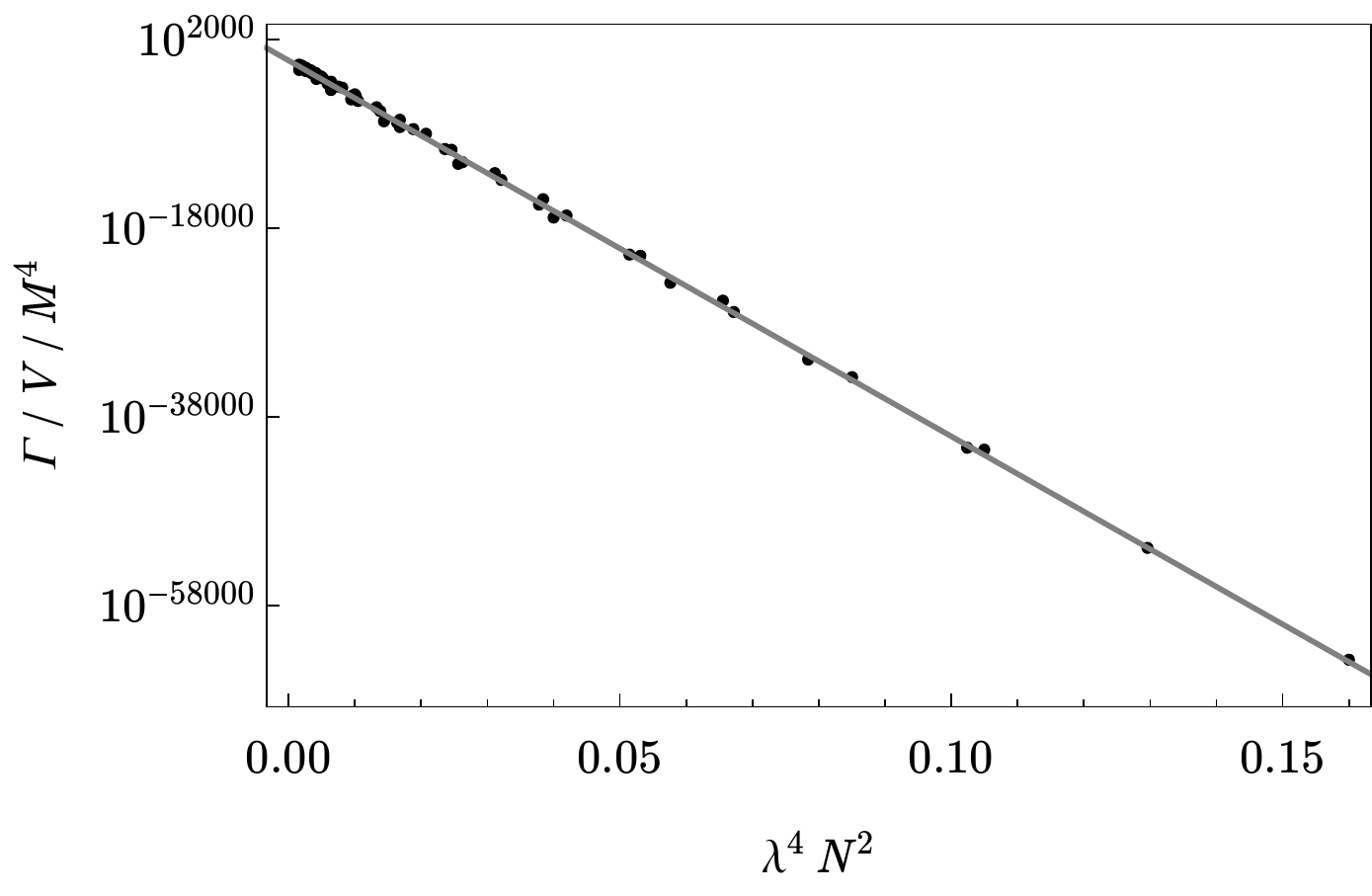}
\caption{\label{fig:rate}Plot of the tunneling rate per unit volume as a function of $\lambda^4N^2$ for $g=0.001$.}
\end{figure}

Finally, in Fig.~\ref{fig:rate}, we plot the tunneling rate as a function of $\lambda^4N^2$ for $g=0.001$, illustrating the severity of the dependence on the parameters in the thin-wall regime, with a variation of $60,000$ orders of magnitude across the factor of $10$ in the expansion parameter $\lambda^2 N$.

Before concluding, we should compare the present approach with the one outlined in Ref.~\cite{Weinberg:1992ds}. Were we to follow the latter work, we would integrate out the fields $X_i$ within a background of homogeneous $\Phi$ configurations. At one loop, this would lead to the renormalized effective potential $U^R_{\rm eff}$ appearing in Eq.~\eqref{eq:pseudo1}. It is therefore clear that the method of Ref.~\cite{Weinberg:1992ds} ignores the small gradient corrections that we have isolated in this section. Besides accounting for gradients in situations where these are more sizable, our present approach will also prove useful in cases where the tree-level potential is non-convex and the effective potential is ill defined, as is indicated by the occurrence of imaginary parts in the loop integrals.

\section{Conclusions}
\label{sec:conc}

We have developed a method for calculating the self-consistent tunneling configuration and one-loop tunneling action for cases in which the global minimum of the potential is generated radiatively. By basing this method upon the 2PI effective action, we are able to deal with the radiatively-induced negative-semi-definite eigenmodes of the one-loop fluctuation operator. Within the context of an $N$ field model with SSB via the Coleman-Weinberg mechanism, we have shown that the incorporation of gradient effects leads only to minor corrections compared to approximate calculations based on the Coleman-Weinberg effective potential of the homogeneous field configuration. However, through an explicit calculation, we have confirmed that the impact of these gradients on the one-loop fluctuation determinant may nevertheless compete with two-loop effects, as has been anticipated previously~\cite{Weinberg:1992ds}.

For the present model, the gradient effects are suppressed as a result of being in the thin-wall regime, wherein the profile of the bubble wall is symmetric about the mid-point of the bounce. The latter means that the bounce itself is going to zero in the region where the gradients are of most relevance [cf. Eqs.~(\ref{eq:pseudo1}) and~(\ref{eq:pseudo2})]. It is anticipated that such suppression will be lessened for models where the thin-wall approximation does not hold and the bounce profile is no longer symmetric.

A pertinent example where gradients may be of decisive importance
is the Fubini-Lipatov instanton~\cite{Fubini:1976jm,Lipatov:1976ny}, 
occurring in the abyssal and conformally-invariant $\lambda\varphi^4$ potential with $\lambda<0$ and of relevance to studies of the stability of the electroweak vacuum of the Standard Model. In the case that the loop corrections to the scalar potential are dominated by
fermions, one easily finds that there are no bounce solutions within the effective potential for homogeneous field backgrounds provided one insists that $|{\varphi}|_{r\,=\,0}<\infty$. It is an interesting question whether this situation changes once gradients are accounted for in determining the self-consistent soliton at one-loop level. In order to decide this
matter, we anticipate that it is straightforward to extend the method presented here beyond the thin- and planar-wall approximations, thereby correctly capturing all leading gradient effects that may occur for tunneling processes between strongly non-degenerate vacua.

\begin{acknowledgments}

The authors would like to thank Holger Gies for helpful discussions. The work  of P.M.~was supported in part by the Science and Technologies Facilities Council (STFC) under Grant No.~ST/L000393/1 and a University  Foundation Fellowship
(TUFF) from  the Technische  Universit\"{a}t M\"{u}nchen. The  work of
B.G. is  supported by the  Gottfried Wilhelm Leibniz programme  of the
Deutsche  Forschungsgemeinschaft  (DFG).    Both  authors  acknowledge
support from the DFG cluster of excellence Origin and Structure of the
Universe.

\end{acknowledgments}

\appendix

\section{Coleman-Weinberg effective potential}
\label{app:CW}

Assuming a homogeneous background, the renormalized one-loop Coleman-Weinberg effective potential takes the form
\begin{equation}
U^R_{\mathrm{eff}} = U+\delta U+\frac{1}{2}\int\!\frac{\D^4 k}{(2\pi)^4}\;\ln\,\mathrm{det}_{ij}\,\bm{G}^{-1}(k)\;,
\end{equation}
where
\begin{align}
\delta U\ &=\ \frac{1}{2}\,\delta m_{\varphi}^2\varphi^2+\frac{1}{2}\,\delta m_{\chi}^2\sum_{i\,=\,1}^N\chi_i^2 +\frac{1}{4}\,\delta\lambda\,\varphi^2\sum_{i\,=\,1}^N\chi_i^2\nonumber\\&\qquad+\frac{1}{4}\sum_{i,j\,=\,1}^N\delta\kappa\chi_i^2\chi_j^2+\frac{1}{4!}\,\delta\alpha\, \varphi^4\;.
\end{align}

In order to calculate the determinant in field space, we require the  eigenvalues $\{m^2\}$ of the mass matrix
\begin{equation}
\bm{m}^2 = \begin{bmatrix} m_{\varphi}^2 & \lambda\varphi\bm{\chi}^{\mathsf{T}} \\ \lambda\varphi\bm{\chi} & \bm{m}^2_{\chi}\end{bmatrix}\;, 
\end{equation}
where
\begin{equation}
\bm{m}^2_{\chi} = \begin{bmatrix} m_{\chi}^2+2\kappa\chi_1^2 & 2\kappa\chi_1\chi_2 & \dots & 2\kappa\chi_1\chi_N \\ 2\kappa\chi_2\chi_1 & m_{\chi}^2+2\kappa\chi_2^2 & \dots & 2\kappa\chi_2\chi_N \\ \vdots & & \ddots & \vdots\\ 2\kappa\chi_N\chi_1 & 2\kappa\chi_N\chi_2 & \dots & m_{\chi}^2+2\kappa\chi_N^2\end{bmatrix}\;,
\end{equation}
and we have defined
\begin{align}
m_{\varphi}^2 =  \frac{\lambda}{2}\sum_{i\,=\,1}^N\chi_i^2\;,\qquad 
m_{\chi}^2 = \kappa\sum_{i\,=\,1}^{N}\chi_i^2+\frac{\lambda}{2}\varphi^2\;.
\end{align}

In block form, the characteristic equation may be written as follows:
\begin{align}
&\|\bm{m}^2-\bm{1}_{N+1}\,m^2\|\ =\ \|\bm{m}_{\chi}^2-\bm{1}_{N}\,m^2 \| \nonumber\\&\qquad \times\big(m_{\varphi}^2-m^2-\lambda^2\varphi^2\bm{\chi}^{\mathsf{T}}\big[\bm{m}^{2}_{\chi}-\bm{1}_{N}\,m^2\big]^{-1}\bm{\chi}\big)\;.
\end{align}
Decomposing the matrix
\begin{equation}
\bm{m}_{\chi}^2\ =\ \bm{D}+2\kappa\bm{\chi}\bm{\chi}^{\mathsf{T}}\;,\qquad \big[\bm{D}\big]_{ij}\ =\ \big(m_{\chi}^2-m^2\big)\delta_{ij}\;,
\end{equation}
it follows from the matrix determinant lemma that
\begin{equation}
\|\bm{D}+2\kappa\bm{\chi}\bm{\chi}^{\mathsf{T}}\|\ =\  \|\bm{D}\|\big(1+2\kappa\bm{\chi}^{\mathsf{T}}\bm{D}^{-1}\bm{\chi}\big)\;,
\end{equation}
in which $\|\bm{D}\|\ =\ \big(m_{\chi}^2-m^2\big)^N$. Thus, we have
\begin{equation}
\|\bm{m}^2_{\chi}-\bm{1}\,m^2\|\ =\ \Bigg[1+2\sum_{k\,=\,1}^N\frac{\kappa\chi_k^2}{m_{\chi}^2-m^2}\Bigg]\big(m_{\chi}^2-m^2\big)^N\;.
\end{equation}

In order to calculate the inverse of $\bm{D}+2\kappa\bm{\chi}\bm{\chi}^{\mathsf{T}}$, we can use the Sherman-Morrison formula, giving
\begin{align}
&\big[\bm{D}+2\kappa\bm{\chi}\bm{\chi}^{\mathsf{T}}\big]^{-1}
 = \bm{D}^{-1}-2\,\frac{\bm{D}^{-1}
\kappa\bm{\chi}\bm{\chi}^{\mathsf{T}}\bm{D}^{-1}}{1+2\kappa\bm{\chi}^{\mathsf{T}}\bm{D}^{-1}\bm{\chi}}\nonumber\\&= \frac{\delta_{ij}}{m_{\chi}^2-m^2} -2\Bigg[1+2\sum_{k\,=\,1}^N\frac{\kappa\chi_k^2}{m_{\chi}^2-m^2}\Bigg]^{\!-1}\frac{\kappa\chi_i\chi_j}{\big(m_{\chi}^2-m^2\big)^2}\;.
\end{align}
Finally, we obtain
\begin{align}
&\big(m_{\chi}^2-m^2\big)^{N-1}\Bigg[\big(m_{\varphi}^2-m^2\big)\big(m_{\chi}^2-m^2\big)\nonumber\\&\qquad+2\bigg(m_{\varphi}^2-m^2-\frac{\lambda^2}{2\kappa}\,\varphi^2\bigg)\sum_{i\,=\,1}^N\kappa\chi_i^2\Bigg]\ =\  0\;,
\end{align}
giving $N-1$ degenerate eigenvalues $m_{\chi}^2$ and two non-degenerate eigenvalues
\begin{align}
& m^2_{\pm} = \frac{m_{\varphi}^2+m_{\chi}^2+2\kappa\sum_{i\,=\,1}^N\chi_i^2}{2}\nonumber\\&\quad \pm\Bigg[\Bigg(\frac{m_{\varphi}^2-m_{\chi}^2-2\kappa\sum_{i\,=\,1}^N\chi_i^2}{2}\Bigg)^2+\lambda^2\varphi^2\sum_{i\,=\,1}^N\chi_i^2\Bigg]^{1/2}\;.
\end{align}

We choose the following renormalization conditions:
\begin{subequations}
\begin{gather}
\frac{\partial^2 U_{\mathrm{eff}}}{\partial\varphi^2}\bigg|_{\varphi\,=\,\chi_i\,=\,0} = 0\;,\qquad
\frac{\partial^2 U_{\mathrm{eff}}}{\partial\chi_i^2}\bigg|_{\varphi\,=\,\chi_i\,=\,0} = 0\;,\\
\frac{\partial^4 U_{\mathrm{eff}}}{\partial\varphi^4}\bigg|_{\varphi\,=\,0,\ \chi_1\,=\,M}\ =\ 0\;,\\
\frac{\partial^4 U_{\mathrm{eff}}}{\partial\varphi^2\partial\chi_i^2}\bigg|_{\varphi\,=\,0,\ \chi_1\,=\,M}\ =\ \lambda\;,\\
\frac{\partial^4 U_{\mathrm{eff}}}{\partial\chi_i^4}\bigg|_{\varphi\,=\,0,\ \chi_i\,=\,M}\ =\ 6\kappa\;,
\end{gather}
\end{subequations}
where the finite scale $M$ is necessary due to the IR singularity of the effective four-point vertices.

The one-loop contributions to the effective potential take the form
\begin{align}
& U_{\mathrm{eff}} \supset \frac{1}{16\pi^2}\bigg\{\Lambda^2\bigg[N m_{\chi}^2+m_{\varphi}^2+2\kappa\sum_{i\,=\,1}^N\chi_i^2\bigg]\nonumber\\&\quad +\frac{1}{4}\bigg[\big(N-1\big)m_{\chi}^4\bigg(\!\ln\frac{m_{\chi}^2}{4\Lambda^2}+\frac{1}{2}\bigg)\nonumber\\&\quad +m_+^4\bigg(\!\ln\frac{m_{+}^2}{4\Lambda^2}+\frac{1}{2}\bigg)+m_-^4\bigg(\!\ln\frac{m_{-}^2}{4\Lambda^2}+\frac{1}{2}\bigg)\bigg]\bigg\}\;,
\end{align}
giving the counterterms
\begin{subequations}
\begin{align}
\delta m_{\varphi}^2\ &= \ -\,\frac{\lambda N}{16\pi^2}\,\Lambda^2\;,\\
\delta m_{\chi}^2\ &=\ -\,\frac{1}{16\pi^2}\big[\lambda+2\big(N+2\big)\kappa\big]\Lambda^2\;,\\
\delta\lambda\ &=\ -\,\frac{\lambda}{16\pi^2}\bigg[\big(3\kappa+2\lambda\big)\ln 3
+\frac{2\lambda^2}{6\kappa-\lambda}\,\ln\frac{6\kappa}{\lambda}\nonumber\\&\hspace{-1em}+\big[2\lambda+\big(N+2\big)\kappa\big]\bigg(\!\ln\frac{\kappa M^2}{4\Lambda^2}+4\bigg)\bigg]\;,\\
\delta\kappa\ &=\ -\,\frac{1}{16\pi^2}\bigg[9\kappa^2\ln3+\frac{\lambda^2}{4}\bigg(\!\ln\frac{\lambda M^2}{8\Lambda^2}+\frac{14}{3}\bigg)\nonumber\\&\hspace{-1em} +\big(N+8\big)\kappa^2\bigg(\ln\frac{\kappa M^2}{4\Lambda^2}+\frac{14}{3}\bigg)\bigg]\;,\\
\delta\alpha\ &=\ -\,\frac{3\lambda^2}{32\pi^2}\bigg[\ln 3+N\bigg(\!\!\ln\frac{\kappa M^2}{4\Lambda^2}+2\bigg)\nonumber\\&\hspace{-1em} +\frac{8\lambda}{\big(6\kappa-\lambda\big)^2}\bigg(6\kappa+3\lambda-\lambda\,\frac{18\kappa+\lambda}{6\kappa-\lambda}\,\ln\frac{6\kappa}{\lambda}\bigg)\bigg]\;.
\end{align}
\end{subequations}

\section{Fluctuation Determinant}
\label{app:Baacke}

In this appendix, we outline the method due to Baacke and Junker~\cite{Baacke:1993jr,Baacke:1993aj,Baacke:1994ix} (see also Refs.~\cite{Baacke:1994bk,Baacke:2008zx}) for calculating the fluctuation determinant in terms of direct integration of the Green's function.

The normalized fluctuation determinant is
\begin{equation}
B^{(1)}[\varphi]\ =\ \frac{N}{2}\Big(\ln\,\mathrm{det}\,S^{-1}(\varphi)\:-\:\ln\,\mathrm{det}\,S^{-1}(v)\Big)\;,
\end{equation}
where the fluctuation operator $S^{-1}(\varphi)$ corresponds to the Klein-Gordon operator, having the form
\begin{equation}
S^{-1}(\varphi)\ =\ -\,\Delta^{(4)}\:+\:m^2(\varphi)\;,
\end{equation}
where $\Delta^{(4)}$ is the four-dimensional Laplace-Beltrami operator and, in our case, $m^2(\varphi)=\lambda\varphi^2/2$.

In the case of spherically-symmetric potentials, it is convenient to work in four-dimensional hyperspherical coordinates, writing $\mathbf{x}=r\mathbf{e}_r$. The eigenfunctions of the fluctuation operator $f_{nj\{\ell\},x}$ may then be expressed via the partial-wave decomposition
\begin{equation}
f_{nj\{\ell\},x}\ =\ \phi_{nj,r}Y_{j\{\ell\},\mathbf{e}_r}\;,
\end{equation}
where  $Y_{j\{\ell\},\mathbf{e}_r}$ are the hyperspherical harmonics (see e.g.~Ref.~\cite{Avery}), and $n$ and $j$, $\{\ell\}=\{\ell_1,\ell_2\}$ are the radial and angular-momentum quantum numbers. The radial functions $\phi_{nj,r}$ satisfy the eigenvalue equation
\begin{equation}
\bigg[-\:\frac{\mathrm{d}^2}{\mathrm{d}r^2}\:-\:\frac{3}{r}\,\frac{\mathrm{d}}{\mathrm{d}r}\:+\:\frac{j(j+2)}{r^2}\:+\:m^2(\varphi)\bigg]\phi_{nj,r}\ =\ \lambda_{nj}\phi_{nj,r}\;.
\end{equation}

In terms of the eigenvalues $\lambda_{nj}$, the normalized fluctuation determinant may be written as
\begin{equation}
B^{(1)}\ = \ \frac{N}{2}\sum_{n,j,\{\ell\}}\ln\frac{\lambda_{nj}}{\lambda^{(v)}_{nj}}\;,
\end{equation}
where the $\lambda_{nj}^{(v)}$ are the eigenvalues of the fluctuation operator in the false vacuum. Note that the fluctuation determinant is formally UV divergent, and it is necessary to regularize the sum over the eigenvalues. The quantum numbers $\{\ell\}$ label the irreducible representations of $SO(4)$, of which there are $(j+1)^2$. The eigenvalues of these representations are degenerate, and we therefore find
\begin{equation}
\label{flucdeteigen}
B^{(1)}\ = \ \frac{N}{2}\sum_{n,j}(j+1)^2\ln\frac{\lambda_{nj}}{\lambda^{(v)}_{nj}}\;.
\end{equation}

In order to obtain an expression for the fluctuation determinant in terms of the inverse of the fluctuation operator, viz.~the Green's function, we consider the operator
\begin{equation}
S^{-1}(\varphi,s)\ \equiv\ S^{-1}(\varphi)\:+\:s\;,
\end{equation}
where $s\in\mathbb{R}$ is an auxiliary parameter. The Green's function can be written as
\begin{equation}
S_{xx'}(\varphi,s)\ =\ \sum_{n,j,\{\ell\}}\frac{f^*_{nj,\{\ell\},x'}f_{nj,\{\ell\},x}}{\lambda_{nj}+s}\;.
\end{equation}
Making use of the sum rule
\begin{equation}
\sum_{\{\ell\}}Y_{j,\{\ell\},\mathbf{e}_r'}^*Y_{j,\{\ell\},\mathbf{e}_r}\ =\ \frac{1}{2\pi^2}\,(j+1)U_j(\cos\theta)\;,
\end{equation}
where $U_j(z)$ are the Chebyshev polynomials of the second kind and $\cos\theta\ =\ \mathbf{e}_r\cdot\mathbf{e}_{r'}$, we find
\begin{equation}
S_{xx'}(\varphi,s)\ =\ \frac{1}{2\pi^2}\sum_{n,j}(j+1)U_j(\cos\theta)\,\frac{\phi^*_{nj,r'}\phi_{nj,r}}{\lambda_{nj}+s}\;.
\end{equation}
Thus, at coincidence ($x=x'$), we obtain
\begin{equation}
S_{xx}(\varphi,s)\ =\ \frac{1}{2\pi^2}\sum_{n,j}(j+1)^2\frac{\phi^*_{nj,r}\phi_{nj,r}}{\lambda_{nj}+s}\;,
\end{equation}
where we have used the fact that $U_j(1)=j+1$.

Integrating $S_{xx}(\varphi,s)$ over $x$, we obtain
\begin{equation}
\int\!\mathrm{d}^4x\; S_{xx}(\varphi,s)\ =\ \sum_{n,j}\frac{(j+1)^2}{\lambda_{nj}+s}
\end{equation}
by virtue of the orthonormality of the radial eigenfunctions,
\begin{equation}
\int_0^{\infty}\!\mathrm{d}r\;r^3\phi^*_{nj,r}\phi_{nj,r}\ =\ 1\;.
\end{equation}
Subsequently, we integrate over $s$ up to some UV cutoff $\Lambda^2$, giving
\begin{equation}
\int_{0}^{\Lambda^2}\!\!\mathrm{d}s \int\!\mathrm{d}^4x\; S_{xx}(\varphi,s)\ =\ -\:\sum_{n,j}(j+1)^2\ln\frac{\lambda_{nj}}{\lambda_{nj}+\Lambda^2}\;.
\end{equation}
Comparing this with Eq.~\eqref{flucdeteigen}, we see by inspection that
\begin{equation}
B^{(1)}[\varphi]\ =\ -\:\frac{N}{2}\int_{0}^{\Lambda^2}\!\!\mathrm{d}s\int\!\mathrm{d}^4x\;\Big(S_{xx}(\varphi,s)\:-\:S_{xx}(v,s)\Big)
\end{equation}
up to UV-divergent terms in the cutoff $\Lambda$. The latter are removed by the addition of the normalized counterterm in Eq.~\eqref{B1counter}.

\end{document}